\documentclass[10pt]{article}

\usepackage{amssymb,amsmath,natbib,textcomp,fullpage,url,multirow,graphicx,parcolumns,color}

\usepackage{kmath,kerkis} % The order of the packages matters; kmath changes the default text font
\usepackage[T1]{fontenc}

\usepackage{tikz}

\usepackage{multicol,float}
\usepackage{titlesec}
\titleformat*{\section}{\large}
\titleformat*{\subsection}{\normalsize\itshape}
\titleformat*{\subsubsection}{\normalsize}

\usepackage[colorlinks=true,citecolor=blue,urlcolor=blue]{hyperref}
\bibpunct{(}{)}{;}{a}{}{,}

\begin{document}

\definecolor{grey}{rgb}{0.9,0.9,0.9}

\pagestyle{myheadings}

\thispagestyle{empty}

\begin{center}
\vspace*{-35pt}
\begin{tikzpicture}
\def\dy{2.5};
\draw [grey,fill=grey] (0,0) rectangle (\textwidth,\dy);
\node [above] at (0.5\textwidth,0.5*\dy) {\huge \textcolor{white}{J}ournal of \textcolor{white}{Q}uantitative \textcolor{white}{S}pectroscopy \&};
\node [below] at (0.5\textwidth,0.5*\dy) {\huge \textcolor{white}{R}adiative \textcolor{white}{T}ransfer};
\end{tikzpicture}
\end{center}

\vspace*{-15pt}
\noindent\rule{\textwidth}{3pt}
~\\
\LARGE
The influence of frequency-dependent radiative transfer on the\\structures of radiative shocks\\
\Large
~\\N. Vaytet\textsuperscript{1,}\footnote{Corresponding author. \textit{E-mail address:} neil.vaytet@ens-lyon.fr}, M. Gonz\'{a}lez\textsuperscript{2}, E. Audit\textsuperscript{3,4}, G. Chabrier\textsuperscript{1,5}\\~\\
\small
\textsuperscript{1}\textit{\'{E}cole Normale Sup\'{e}rieure de Lyon, CRAL, UMR CNRS 5574, Universit\'{e} de Lyon, 46 All\'{e}e d'Italie, 69364 Lyon Cedex 07, France}\\
\textsuperscript{2}\textit{Universit\'{e} Paris Diderot, Sorbonne Paris Cit\'{e}, AIM, UMR 7158, CEA, CNRS, F-91191 Gif-sur-Yvette, France}\\
\textsuperscript{3}\textit{Maison de la Simulation, USR 3441,  CEA - CNRS - INRIA - Universit\'{e} Paris-Sud - Universit\'{e} de Versailles, F-91191 Gif-sur-Yvette, France}\\
\textsuperscript{4}\textit{CEA/DSM/IRFU, Service d'Astrophysique, Laboratoire AIM, CNRS, Universit\'{e} Paris Diderot, F-91191 Gif-sur-Yvette, France}\\
\textsuperscript{5}\textit{School of Physics, University of Exeter, Exeter, EX4 4QL, UK}\\
\normalsize
\hrule
~\\
\begin{parcolumns}[nofirstindent,colwidths={1=.25\linewidth}]{2}
\colchunk{A~R~T~I~C~L~E~~~I~N~F~O\\ \line(1,0){135}\\
\textit{Article history:}\\ Received 7 November 2012\\ Accepted 6 March 2013\\ Available online 7 March 2013\\
\line(1,0){135}Moment model\\ Multigroup\\ Laboratory astrophysics\\ Numerical methods\\ Shock waves\\
\\~\\~\\~\\~\\~\\~\\~\\~\\~\\~\\~\\~\\ \line(1,0){135}\\}
\colchunk{A~B~S~T~R~A~C~T\\ \line(1,0){388.5}\\ 
Radiative shocks are shocks in a gas where the radiative energy and flux coming from the very hot post-shock
material are non-negligible in the shock's total energy budget, and are often large enough to heat the material ahead of the
shock. Many simulations of radiative shocks, both in the contexts of astrophysics and laboratory experiments, use a grey
treatment of radiative transfer coupled to the hydrodynamics. However, the opacities of the gas show large variations as a
function of frequency and this needs to be taken into account if one wishes to reproduce the relevant physics. We have
performed radiation hydrodynamics simulations of radiative shocks in Ar using multigroup (frequency dependent) radiative
transfer with the \textsc{heracles} code. The opacities were taken from the \textsc{odalisc} database. We show the influence
of the number of frequency groups used on the dynamics and morphologies of subcritical and supercritical radiative shocks in
Ar gas, and in particular on the extent of the radiative precursor. We find that simulations with even a low number of groups
show significant differences compared to single-group (grey) simulations, and that in order to correctly model such shocks, a
minimum number of groups is required. Results appear to eventually converge as the number of groups increases above 50. We
were also able to resolve in our simulations of supercritical shocks the adaptation zones which connect the cooling layer to
the final post-shock state and the precursor. Inside these adaptation zones, we find that the radiative flux just ahead of the
shock in one or several high-opacity groups can heat the gas to a temperature higher than the post-shock temperature. Through
the use of Hugoniot curves, we have checked the consistency of our radiation hydrodynamics scheme by showing that conservation
of mass, momentum and energy (including radiative flux) holds throughout the computational domain for all our simulations. We
conclude that a minimum number of frequency groups are required to correctly simulate radiating flows in gases whose opacities
present large variations as a function of frequency.\\ \line(1,0){388.5}\\}
\end{parcolumns}

\begin{multicols}{2}

\section{Introduction}\label{sec:introduction}

Radiative shocks are shocks in a gas where the radiative energy and flux coming from the very hot post-shock material are non-
negligible in the shock's total energy budget \citep{zeldovich1967,mihalas1984}. A radiative precursor is formed ahead of the
shock when the forward flux of ionizing photons exceeds the flux of atoms approaching the shock front. These conditions are met
when the shock velocity exceeds the threshold required to produce the necessary photon flux \citep{keiter2002}. Two regimes of
radiative shocks are often described in the literature. The first is called the subcritical regime, where the shock has only a
transmissive precursor and the temperature just ahead of the discontinuity is not equal to the final downstream state temperature.
The second is known as the supercritical regime which arises as the strength of the shock increases; a diffusive region in the
precursor appears and the pre-shock temperature reaches the final state temperature \citep[see][for more details]{drake2006}.

\begin{figure}[H]
\centering
\includegraphics[scale=0.65]{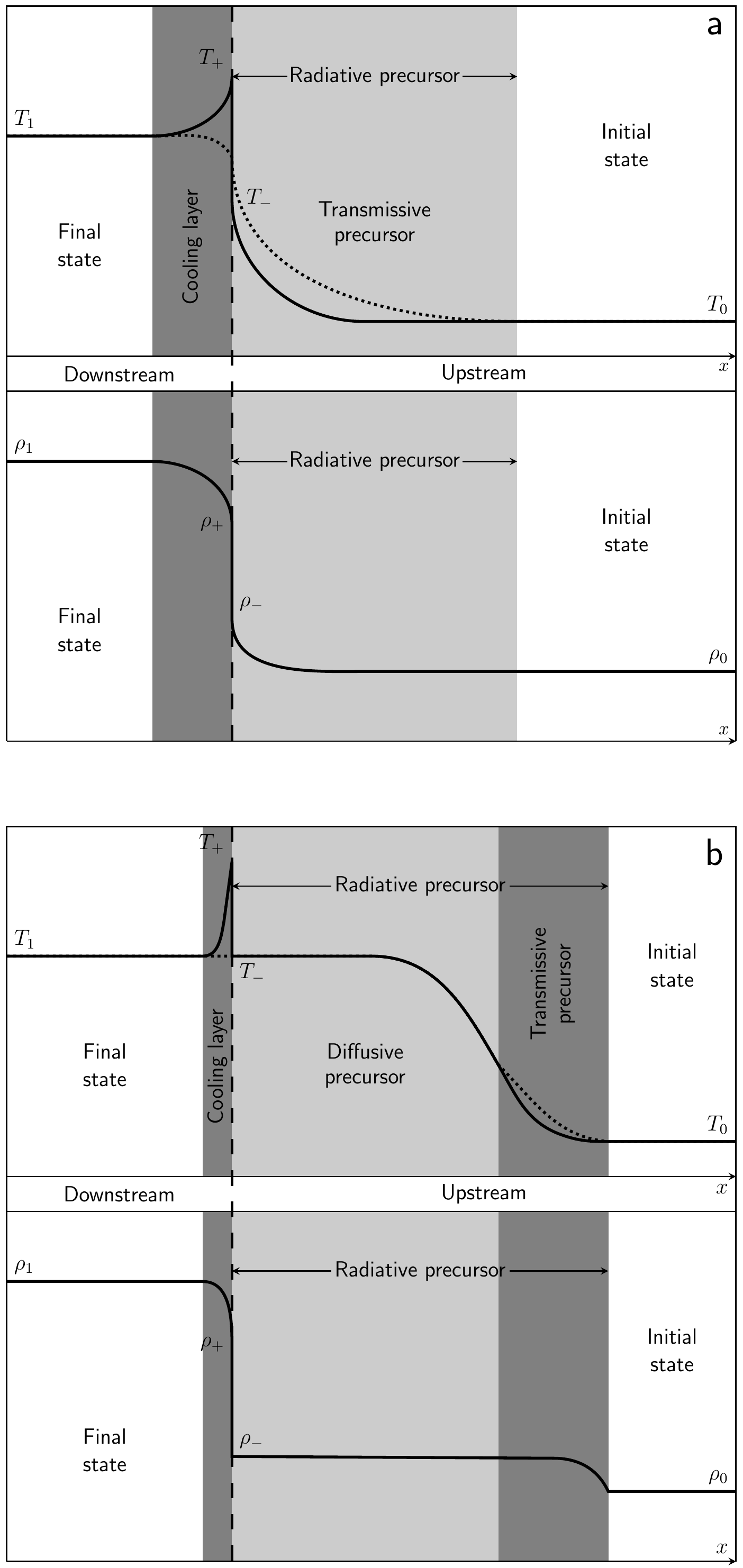}
\caption{Classical structure of a subcritical (a) and supercritical (b) radiative shock. The direction of the gas flow is from right to
left in the frame where the shock is at rest. The panels show the gas (solid) and radiative (dotted) temperature (top) and the gas density
(bottom) as a function of distance in each case. The position of the temperature and density jumps is marked by the vertical dashed line.
The relative sizes of the layers are for illustration purposes only.}
\label{fig:sub_and_sup_shock_structure}
\end{figure}

The classical structure of a subcritical radiative shock is depicted in Fig.~\ref{fig:sub_and_sup_shock_structure}a. The preshock gas
is heated by the radiative precursor to a temperature $T_{-}$ and the shock compression heats it further to a temperature $T_{+}$
which is higher than the final post-shock equilibrium state $T_{1}$. The gas then cools down inside the cooling layer to reach the
equilibrium state $T_{1}$ by radiating the excess energy away. The radiation is decoupled from the gas inside the cooling layer and
the transmissive precursor. The pressure gradient in the precursor, through the conservation of mass and momentum, causes the
velocity to decrease and the density to increase to a value $\rho_{-}$ ahead of the discontinuity. The sharp density
jump of the shock from $\rho_{-}$ to $\rho_{+}$ then takes place on the gas viscous scale. The density increases further from
$\rho_{+}$ to $\rho_{1}$ inside the cooling layer as the gas contracts \citep[see also][]{zeldovich1967,drake2006}.

In the case of an optically thick supercritical radiative shock (shown in Fig.~\ref{fig:sub_and_sup_shock_structure}b), we have
$T_{-} \approx T_{1}$. The radiative temperature is equal to the gas temperature for the most part, except that it remains constant
across the cooling layer and is higher than the gas temperature inside the transmissive precursor \citep{mihalas1984}. As stated
above, the pressure gradient at the head of the precursor causes the density to increase. Since the gas temperature is close to
being constant in the diffusive part of the precursor, there is no more pressure gradient and the density reaches a plateau value
$\rho_{-}$ ahead of the discontinuity.

The study of radiative shocks begun in the late 1940s with theoretical studies on the Rankine-Hugoniot jump conditions including a
non-negligible radiation pressure for very energetic shocks \citep[see][for a short review]{blinnikov2011}. The studies were very
rapidly pursued and extended in the field of astrophysics since no processes on Earth could achieve high enough energies to produce
such shocks. Radiative shocks are indeed found in novae \citep{vaytet2007,orlando2009,bode2008}, supernovae
\citep{draine1993,ghavamian2000,nymark2006}, stellar atmospheres \citep{fadeyev2000,gillet2006}, accretion processes in star formation
\citep{stahler1980,commercon2011}, symbiotic stars \citep{falize2009,imamura1985} and jets \citep{raga1999,reipurth1999}. This
omnipresence makes them a key physical process at the heart of high energy astrophysics, and it is thus essential to fully understand
the details of such a mechanism.

In recent years, with the modern advances in technology, radiative shocks have been produced in a number of laboratory laser facilities
\citep[see][for example]{bozier1986,edwards2001,fleury2002,reighard2006,busquet2007,michaut2009}, where very high-energy lasers are used
to drive radiative shocks inside gas chambers. This allows new diagnostics of the properties of
radiative \hfill shocks \hfill with \hfill a \hfill much \hfill more \hfill detailed \hfill view \hfill than

\noindent would ever be possible in astrophysics. Radiative shock experiments allow for the validation of numerical simulations
which are overwhelmingly used in high-energy physics and astrophysics to make predictions on high-energy flows.

Semi-analytic studies of the structure of radiative shocks have been carried out by \citet{lowrie2007,lowrie2008}, and several comparisons
between experiments and simulations have also been undertaken by \citet{bouquet2004,leibrandt2005,reighard2007,gonzalez2009}, for example.
One key piece of data that is required by the numerical simulations in order to accurately model the flow is the opacities of the gas in
which the shock is launched. Gas opacities show large variations as a function of temperature and density as well as frequency, and
including detailed opacities in simulations have crucial effects on the structures of radiative shocks. \citet{vaytet2011} performed
simulations of a radiative shock in Xe using a realistic opacity set and showed the importance of taking into account the frequency
dependence of the opacities, rather than simply integrating over all frequencies, as is commonly done in simulations of radiative shocks

This paper aims to build on the idea that a frequency dependent treatment of radiative transfer is crucial in simulations of radiative
shocks. In particular, we experiment further with the multigroup method of \citet{vaytet2011} by studying the effect of the number of
frequency groups on the structures of the radiative shocks (mainly the variations in size of the precursor). We performed simulations
of stationary radiative shocks (both sub- and supercritical) using 1 to 100 frequency groups, and the differences between the results are 
discussed. The opacities are a crucial part of the radiative transfer model; they govern the amount of energy that will be absorbed and
emitted by the gas and can hence determine the structure of the flow. Laboratory experiments use high atomic number gases to launch
radiative shocks to take advantage of the strong gas heating due to the lower heat capacity. Common choices are argon (Ar) and xenon inert
gases, and in this work we have chosen to use Ar (see section \ref{sec:opacities} for more details).

\section{The multigroup RHD simulations of radiative shocks}\label{sec:simulations}

\subsection{Radiative transfer}\label{sec:radtransfer}

We use the $M_{1}$ moment model \citep{levermore1984,dubroca1999} to approximate the equation of radiative transfer. The $M_{1}$ method
uses the first two moment equations governing the evolution of the radiative energy and flux
\begin{equation}\label{eq:momentequ}
\begin{array}{lcrcl}
\partial_{t} E          &+&     \nabla \cdot \mathbf{F} &=&   \sigma (4\pi B - c E) \\
\partial_{t} \mathbf{F} &+& c^2 \nabla \cdot \mathbb{P} &=& - \sigma c \mathbf{F} 
\end{array}
\end{equation}
where $c$ is the speed of light, $\sigma$ the absorption/emission coefficient and $B$ the black body specific intensity. $E, \mathbf{F}$,
and $\mathbb{P}$ are the zeroth, first and second moments of the radiation specific intensity, namely the radiative energy density, the
radiative energy flux, and the radiative pressure, respectively. In order to close system~(\ref{eq:momentequ}), the radiative pressure is
expressed as a function of the radiative energy and flux following $\mathbb{P} = \mathbb{D} E$ where $\mathbb{D}$ is known as the Eddington
tensor. The expression for $\mathbb{D}$ is obtained by minimizing the radiative entropy which yields
\begin{equation}\label{eq:eddtensor}
\mathbb{D} = \frac{1 - \chi}{2} ~ \mathbb{I} + \frac{3\chi - 1}{2} 
~ \frac{\mathbf{F} \otimes \mathbf{F}}{\|\mathbf{F}\|^{2}}
\end{equation}
where
\begin{equation}\label{eq:chi1}
\chi  = \frac{3 + 4 f^{2}}{5 + 2 \sqrt{4 - 3 f^{2}}}
\end{equation}
and $f = \| \mathbf{F} \| / c E$ is known as the reduced flux. Note that by definition of $E$ and $\mathbf{F}$, we have $f \le 1$, which
implies that  the radiative energy is transported at most at the speed of light. In one dimension we simply have $P = \chi E$. This closure
relation recovers the two asymptotic regimes of radiative transfer. In the free-streaming limit (i.e. transparent media), we have $f = 1$
and $\chi = 1$. On the other hand, in the diffusion limit, $f = 0$ and $\chi=1/3$, which corresponds to an isotropic radiation pressure.

\subsection{The equations of multigroup radiation hydrodynamics}\label{sec:mrhd_equations}

We use the multigroup version of the $M_{1}$ model coupled to the gas hydrodynamics described in \citet{vaytet2011} to account for frequency
dependence of the absorption and emission coefficients \citep[see][for other examples of Godunov multigroup methods]{shestakov2008,vanderholst2011,zhang2013}.
In the multigroup model, the equations of radiative transfer are integrated into a finite number of frequency bins (or groups) and the
opacities are averaged over the same frequency ranges. The closure relation \ref{eq:chi1} is applied within each frequency group. The more the
frequency groups, the more accurate the methods becomes, but the higher the computational cost. The system of multigroup RHD equations in the
comoving frame is
\begin{alignat}{2}
\partial_t \rho             & + \nabla \cdot (\rho \mathbf{u})                                  & & =  0 \label{eq:cons_mass}\\
\partial_t(\rho \mathbf{u}) & + \nabla \cdot (\rho \mathbf{u} \otimes \mathbf{u} + p\mathbb{I}) & & = \sum_{g=1}^{Ng} (\sigma_{Fg}/c) \mathbf{F}_{g} \label{eq:cons_mom}\\
\partial_t e                & + \nabla \cdot (\mathbf{u}(e + p))                                & & = \sum_{g=1}^{Ng} \Big[ c\Big(\sigma_{Eg} E_{g} - \sigma_{Pg} \Theta_{g}(T)\Big)\notag\\
                          ~ &                                                                ~  & & ~~~~ + (\sigma_{Fg}/c) \mathbf{u} \cdot \mathbf{F}_{g} \Big] \label{eq:cons_ener}\\
\partial_{t} E_{g}          & + \nabla \cdot \mathbf{F}_{g} + \nabla \cdot (\mathbf{u} E_{g}) +  \mathbb{P}_{g} & & : \nabla \mathbf{u} \notag\\
- \nabla \mathbf{u}         & : \displaystyle \int_{\nu_{g-1/2}}^{\nu_{g+1/2}}\partial_{\nu}(\nu \mathbb{P}_{\nu}) d\nu  & & = c \big( \sigma_{Pg} \Theta_{g}(T) - \sigma_{Eg} E_{g} \big) \label{eq:cons_Er}\\
\partial_{t} \mathbf{F}_{g} & + c^{2} \nabla \cdot \mathbb{P}_{g} + \nabla \cdot (\mathbf{u} \otimes \mathbf{F}_{g}) & & + \mathbf{F}_{g} \cdot \nabla \mathbf{u} \notag\\
- \nabla \mathbf{u}         &  : \displaystyle \int_{\nu_{g-1/2}}^{\nu_{g+1/2}}\partial_{\nu}(\nu \mathbb{Q}_{\nu}) d\nu  & & = - \sigma_{Fg} c \mathbf{F}_{g} \label{eq:cons_Fr}
\end{alignat}
% \begin{alignat}{1}
% &\partial_{t} \rho             + \nabla \cdot (\rho \mathbf{u})                                  =  0 \label{eq:cons_mass}\\
% &\partial_{t}(\rho \mathbf{u}) + \nabla \cdot (\rho \mathbf{u} \otimes \mathbf{u} + p\mathbb{I}) = \sum_{g=1}^{Ng} (\sigma_{Fg}/c) \mathbf{F}_{g} \label{eq:cons_mom}\\
% &\partial_{t} e                + \nabla \cdot (\mathbf{u}(e + p))                                = \sum_{g=1}^{Ng} \Big[ c\Big(\sigma_{Eg} E_{g} - \sigma_{Pg} \Theta_{g}(T)\Big) + (\sigma_{Fg}/c) \mathbf{u} \cdot \mathbf{F}_{g} \Big] \label{eq:cons_ener}\\
% &\partial_{t} E_{g}            + \nabla \cdot \mathbf{F}_{g} + \nabla \cdot (\mathbf{u} E_{g}) + \mathbb{P}_{g} : \nabla \mathbf{u} - \nabla \mathbf{u} : \displaystyle \int_{\nu_{g-1/2}}^{\nu_{g+1/2}}\partial_{\nu}(\nu \mathbb{P}_{\nu}) d\nu = c \big( \sigma_{Pg} \Theta_{g}(T) - \sigma_{Eg} E_{g} \big) \label{eq:cons_Er}\\
% &\partial_{t} \mathbf{F}_{g}   + c^{2} \nabla \cdot \mathbb{P}_{g} + \nabla \cdot (\mathbf{u} \otimes \mathbf{F}_{g}) + \mathbf{F}_{g} \cdot \nabla \mathbf{u} - \nabla \mathbf{u} : \displaystyle \int_{\nu_{g-1/2}}^{\nu_{g+1/2}}\partial_{\nu}(\nu \mathbb{Q}_{\nu}) d\nu = - \sigma_{Fg} c \mathbf{F}_{g} \label{eq:cons_Fr}
% \end{alignat}
where $c$ is the speed of light, and $\rho$, $\mathbf{u}$, $p$ and $e$ are the gas density, velocity, pressure and total energy,
respectively. $\mathbb{Q}_{\nu}$ is the third moment of the radiation specific intensity; the radiative heat flux. Subscripts $\nu$ denote
monochromatic quantities, and we also define
\begin{equation}\label{eq:groupvar}
X_{g} = \int_{\nu_{g-1/2}}^{\nu_{g+1/2}} X_{\nu} d\nu
\end{equation}
which represents for $X = E$, $\mathbf{F}$, $\mathbb{P}$ the radiative energy, flux and pressure inside each group $g$ which holds
frequencies between $\nu_{g-1/2}$ and $\nu_{g+1/2}$. $N_{g}$ is the total number of groups and $\Theta_{g}(T)$ is the energy of the
photons  having a Planck distribution at temperature $T$ inside a given group. The quantities $\sigma_{Pg}$, $\sigma_{Eg}$ and
$\sigma_{Fg}$ are the means of the absorption/emission coefficient $\sigma_{\nu}$ inside a given group weighted by the Planck function,
the radiative energy and the radiative flux, respectively. The radiative quantities are expressed in the frame comoving with the fluid,
which allows simple expressions to be used for the source terms on the right hand side of equations (\ref{eq:cons_mom})--(\ref{eq:cons_Fr}).
The terms involving the frequency derivatives of the radiative pressure and heat flux $\partial_{\nu}(\nu \mathbb{P}_{\nu})$ and
$\partial_{\nu}(\nu \mathbb{Q}_{\nu})$ are solved using a finite volume method in the frequency dimension \citep[see][for details]{vaytet2011}.

\subsection{Numerical method}\label{sec:numerical_method}

We have implemented the multigroup radiative transfer module of \citet{vaytet2011} in the 3D radiation hydrodynamics second order Godunov
code \textsc{heracles}\footnote{\url{http://irfu.cea.fr/Projets/Site_heracles/}} \citep{gonzalez2007}. It uses an explicit solver for the
hydrodynamics and an implicit solver for the radiative transfer. The Ar gas equation of state is a simple modified ideal gas equation of
state; the ionization energy is neglected but the ionization state is used to compute the mean molecular weight which in turn affects the
gas temperature. The disregard of the ionization energy may greately over-estimate the temperature \citep[ionization can represent half
of the internal energy for mid- to highly ionized flows;][]{drake2006}, but since we focus solely on the differences between mono- and
multi-frequency radiative transfer methods and no comparison between simulations and experiments is made throughout, this oversight does
not matter for the purposes of the present paper.

In the RHD equations, it is not trivial to compute the radiative energy and flux-weighted mean opacities $\sigma_{Eg}$ and $\sigma_{Fg}$.
Common practise is to set $\sigma_{Eg} = \sigma_{Pg}$ and $\sigma_{Fg} = \sigma_{Rg}$ where $\sigma_{R}$ is the Rosseland mean opacity.
In this work, we have used an average opacity $\sigma_{Eg} = \sigma_{Fg} = \sigma_{Ag}$ which varies between the values of $\sigma_{Pg}$
and $\sigma_{Rg}$ depending on the reduced flux $f$ (see Appendix~\ref{sec:appendixA} for more details). However, we wish to point out that the
inaccuracies which arise from these different approximations are reduced as the number of frequency groups used increases, since in the
limit of infinite frequency resolution, all of these quantities simply reduce to $\sigma_{\nu}$. Any approximation is thus less crude in
a multigroup model than in a grey model. The simulations were run on a varying number of CPUs ranging from 12 for the low numbers of
frequency groups to 200 for the heavier calculations. 

\subsection{Initial and boundary conditions}\label{sec:initial_conditions}

The simulations of stationary radiative shocks were performed in a one-dimensional regular cartesian grid comprising 1000 cells (see
Appendix~\ref{sec:appendixB} for a discussion on resolution). The grid sizes were $L = 1$ cm and $L = 6$ cm for the simulations of
subcritical and supercritical radiative shocks, respectively. The discontinuity was initially located at $x_{s} = L/4$

\begin{table}[H]
\begin{center}
\caption[Simulation results]{Simulation results}
\begin{tabular}{@{}c@{~~~}c@{~~~}c@{~~~}c@{~~~}c@{~~~}c@{}}
\hline
\hline
Run             & Number & Shock                & Mach                   & Shock    & Precursor \\
name            & of     & velocity             & number                 & position & size      \\
~               & groups & (km~s$^{-1}$)        & ~                      & (cm)     & (cm)      \\ 
\hline
\texttt{SUB001} & 1      & \multirow{6}{*}{30}  & \multirow{6}{*}{5.62}  & 0.250    & 0.016     \\
\texttt{SUB005} & 5      &                      &                        & 0.250    & 0.025     \\
\texttt{SUB010} & 10     &                      &                        & 0.250    & 0.024     \\
\texttt{SUB020} & 20     &                      &                        & 0.250    & 0.052     \\
\texttt{SUB050} & 50     &                      &                        & 0.250    & 0.070     \\
\texttt{SUB100} & 100    &                      &                        & 0.250    & 0.073     \\
\hline
\texttt{SUP001} & 1      & \multirow{6}{*}{100} & \multirow{6}{*}{18.75} & 1.269    & 2.277     \\
\texttt{SUP005} & 5      &                      &                        & 1.245    & 2.475     \\
\texttt{SUP010} & 10     &                      &                        & 1.239    & 2.582     \\
\texttt{SUP020} & 20     &                      &                        & 1.233    & 2.650     \\
\texttt{SUP050} & 50     &                      &                        & 1.227    & 2.790     \\
\texttt{SUP100} & 100    &                      &                        & 1.227    & 2.811     \\
\hline
\end{tabular}
\label{tab:simulations_results}
\end{center}
{\footnotesize Note: The position of the shock is defined as the position where the derivative of the velocity is the maximum. The size
of the precursor is measured between the shock position and the first point (from the right hand side) where the gas temperature exceeds 1.1
eV.}
\end{table}

\noindent and the gas to the
right of the discontinuity was given a density of $\rho_{0} = 10^{-3}~\text{g cm}^{-3}$ and a temperature of $k_{B}T = 1$ eV. The
radiative temperature was in equilibrium with the gas and the
radiative flux was zero. The upstream velocity was set to $u_{0} = -30~\text{km s}^{-1}$
in the subcritical case and $u_{0} = -100~\text{km s}^{-1}$ for the supercritical shock. Once the upstream state was chosen, the downstream
state was calculated using the Rankine-Hugoniot jump conditions for a radiating fluid \citep{mihalas1984}, which describe the conservation
of mass, momentum and energy across the discontinuity. We find (using the appropriate states of ionisation) the downstream quantities for the
subcritical shock to be $\rho_{1} = 3.65 \times 10^{-3}~\text{g cm}^{-3}$, $u_{1} = -8.21~\text{km s}^{-1}$ and $k_{B}T_{1} = 6.92$ eV. In
the case of the supercritical shock, we obtain $\rho_{1} = 3.97 \times 10^{-3}~\text{g cm}^{-3}$, $u_{1} = -25.18~\text{km s}^{-1}$ and
$k_{B}T_{1} = 32.85$ eV. The upstream and downstream values are also imposed inside ghost cells at the right and left boundaries of the
computational domain, respectively. In computing the upstream and downstream states, we have assumed that we are sufficiently far from the
shock so that the radiative temperature is in equilibrium with the gas temperature and that the radiative flux is zero, which is the case in
our simulations.

\subsection{The Argon opacities and the decomposition of the frequency domain}\label{sec:opacities}

\begin{figure*}[!ht]
\centering
\includegraphics[scale=0.4]{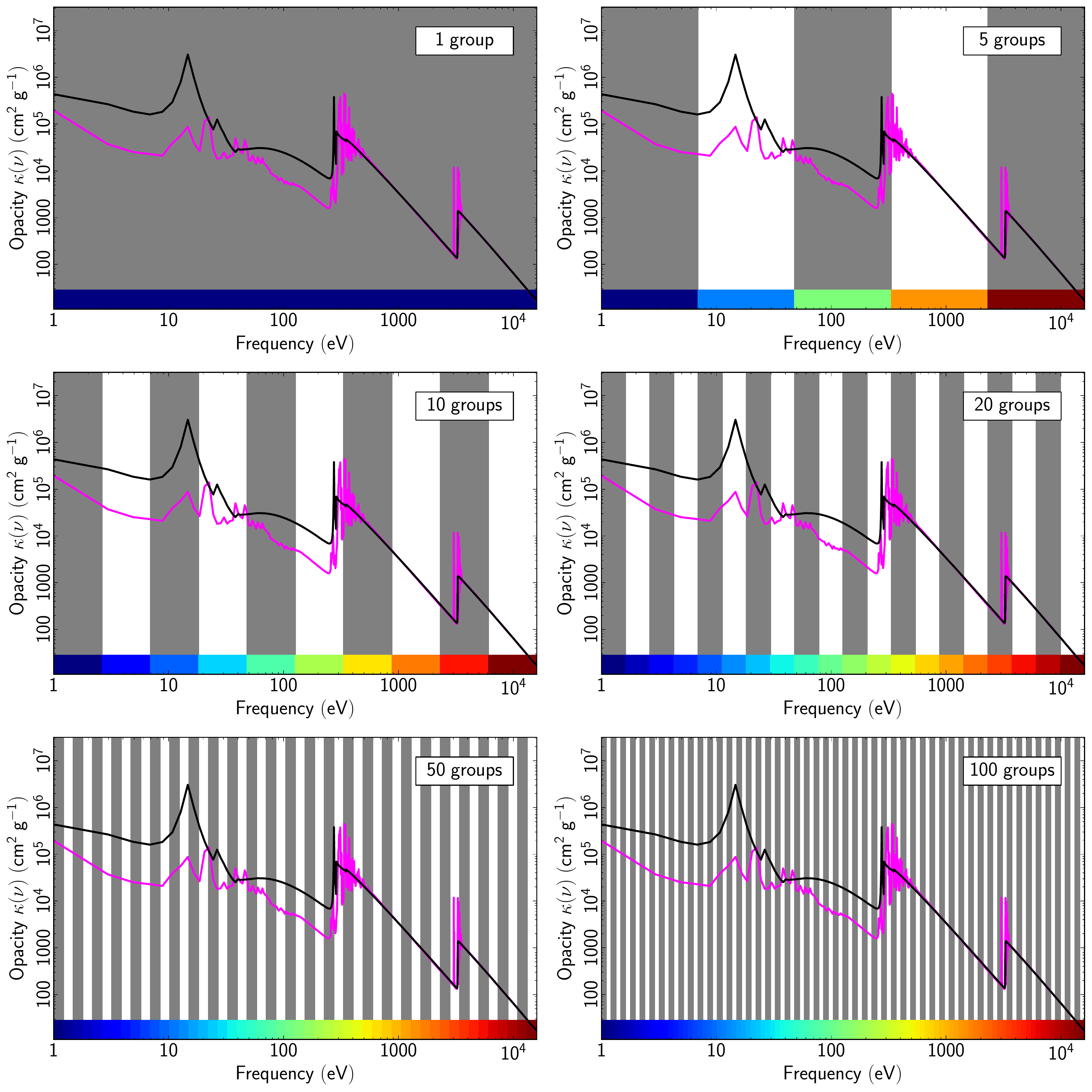}
\caption{The Ar opacities with the decomposition of the frequency domain for 1 to 100 groups. The black line represents the opacities
of the gas in the pre-shock $(\rho,T)$ state while the magenta line is for the post-shock state of the supercritical case. The colour-bar
at the bottom of each frame codes for the group number; this is used in Figs~\ref{fig:argon_sub_shock}, \ref{fig:argon_sup_shock} and
\ref{fig:argon_sup_shock_zoom}.}
\label{fig:kappanugr}
\end{figure*}

The opacities for the Ar gas were taken from the \textsc{odalisc}\footnote{\url{http://irfu.cea.fr/Projets/Odalisc/}}
database, which provides spectral opacities as well as mean opacities (Rosseland and Planck) of many elements for a wide range of physical
conditions. We used the Ar opacities in the frequency range $h\nu = 1 - \textnormal{16,000}$ eV, computed with the \textsc{potrec} code
\citep{mirone1997} which is based on the average atom model, including $l$-splitting and $\Delta n = 0$ transitions. The spectral opacities
for two different densities and temperature are shown in Fig.~\ref{fig:kappanugr}; the pre-shock state is represented by the black line
($\rho_{0} = 10^{-3} \text{g~cm}^{-3} ; k_{B}T_{0} = 1$ eV) while the magenta line is for the supercritical post-shock state ($\rho_{1} \sim 4
\times 10^{-3}~\text{g cm}^{-3} ; k_{B}T_{1} \sim 33$ eV). One can see that the opacities show important orders-of-magnitude variations
as a function of frequency as well as gas density and temperature, and this can have a large impact on the results. A frequency-averaged
model is in this case not an accurate approximation since it cannot model situations where a gas would be optically thick in one part of the
spectrum and optically thin in another.

In this work, we performed simulations with 1, 5, 10, 20, 50 and 100 frequency groups. Choosing the appropriate decomposition of the frequency
domain among the groups is not very straightforward. Ideally, as soon as a large variation in $\kappa_{\nu}$ as a function of frequency is
present, one would require a new frequency group. When only a small number of groups is used, different boundary choices can have different
impacts on the simulation results. Group boundary placing naturally becomes less and less important as the number of frequency groups
increases. For the benefit of a fair comparison between simulation results, the decomposition of the frequency domain among the groups was
simply done logarithmically, as shown in Fig.~\ref{fig:kappanugr}. 

\section{Results}\label{sec:results}

\begin{figure*}[!ht]
\centering
\includegraphics[scale=0.49]{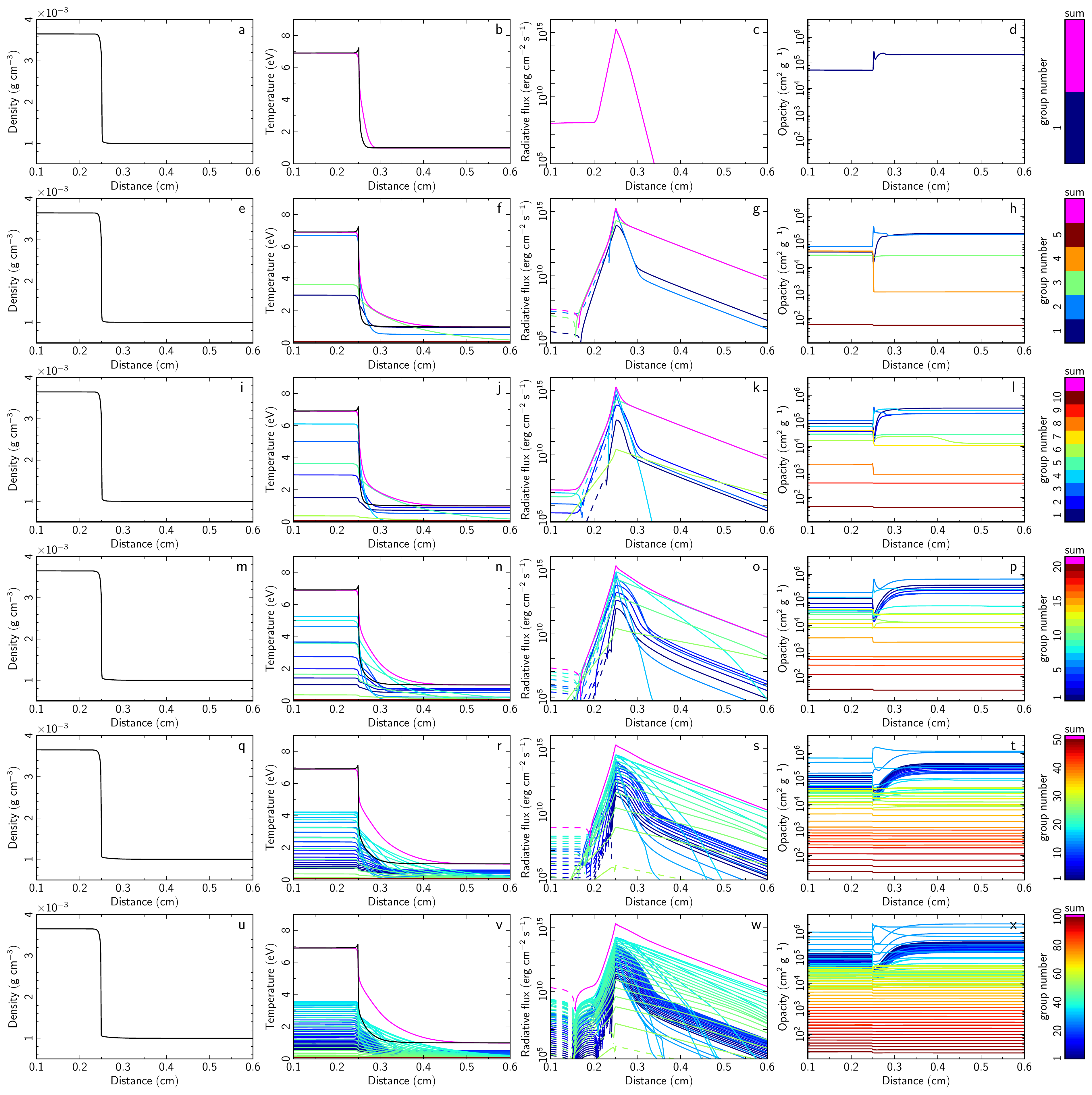}
\caption{Stationary subcritical radiative shocks using (from top to bottom) 1 to 100 frequency groups. From left to right: gas density,
gas (black) and radiative (colours) temperatures, radiative flux and gas opacity. At the end of each row is a colour bar coding for the
different frequency groups. In the two central columns, the magenta curves represent the sum over all groups for the radiative
temperature and flux. Dashed lines represent negative values of the radiative flux (i.e. flowing from right to left).}
\label{fig:argon_sub_shock}
\end{figure*}

We performed simulations of subcritical and supercritical radiative shocks, using $1-100$ frequency groups in both cases. The properties
of the different runs are listed in Table~\ref{tab:simulations_results}. The initial discontinuity, set up with the Rankine-Hugoniot
conditions described in \ref{sec:initial_conditions}, was left to evolve until all the structure in the radiative shock was fully developed
and the stationary regime was reached. All the results shown below (apart from figures showing a time evolution) are in the stationary regime.

\subsection{The subcritical case}\label{sec:sub_shock}

\begin{figure*}[!ht]
\centering
\includegraphics[scale=0.60]{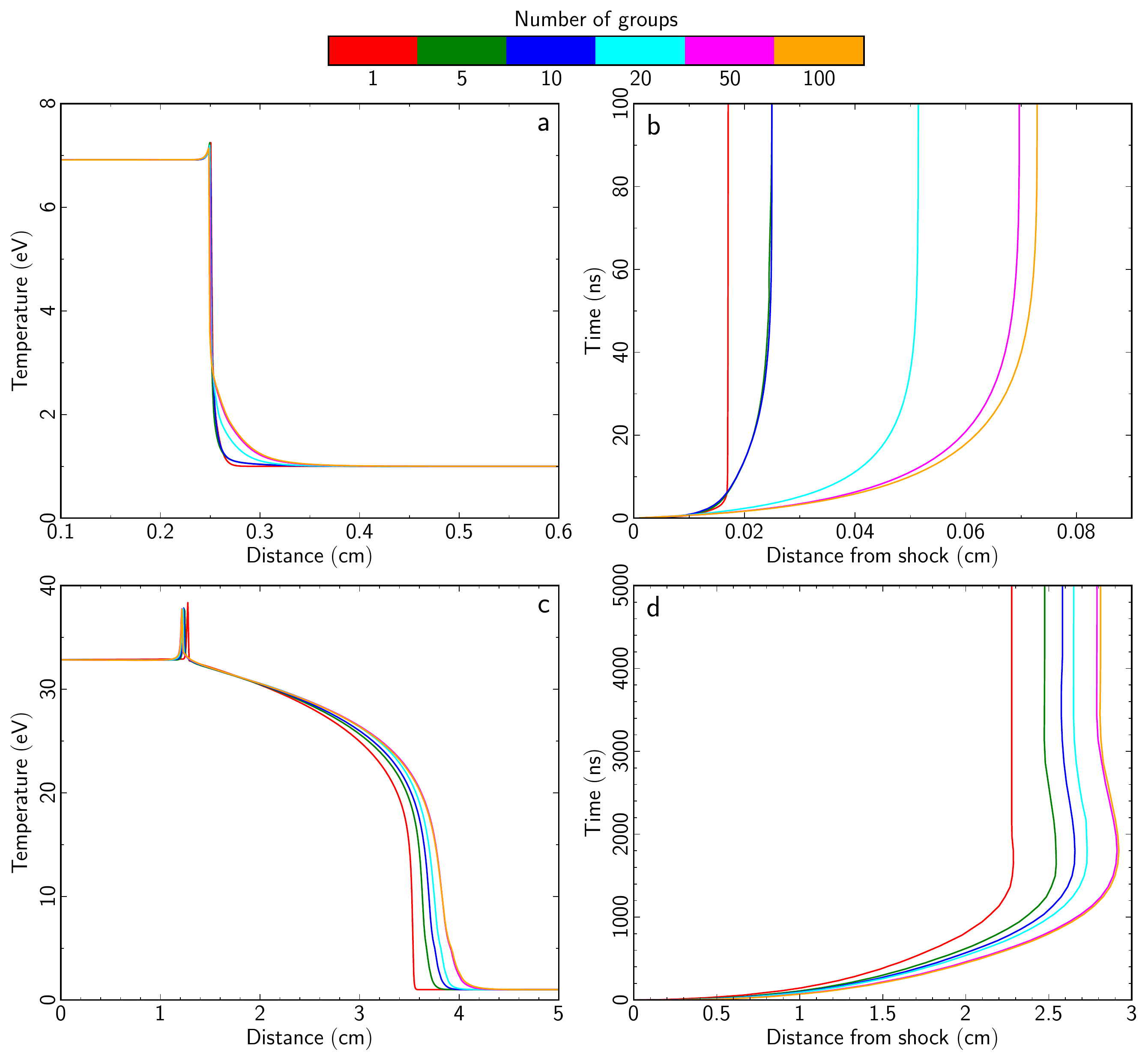}
\caption{Left column: gas temperature as a function of distance using $1-100$ frequency groups (see color legend at the top) in the case of
subcritical (b) and supercritical (d) radiative shocks.
Right column: distance between the head of the radiative precursor and the shock density jump in the case of subcritical (a) and supercritical
(c) radiative shocks.}
\label{fig:precursor_positions_and_temperatures}
\end{figure*}

The results for the simulations of a subcritical radiative shock using $1-100$ frequency groups are shown
in Fig.~\ref{fig:argon_sub_shock}. Each row is for a different number of groups. The columns from left to right display (as a function of
distance) the gas density, the gas (black) and radiative (colours) temperatures, the radiative flux and the gas opacity. In the two middle
columns, the magenta curves represent the sums over all groups.

We first take a look at the mono-group simulation; run \texttt{SUB001} (top row). The profiles exhibit the classic structure
of a subcritical radiative shock; a transmissive radiative precursor extends ahead of the shock heating the upstream gas and altering its
density and velocity (not shown), the radiative flux (c) is maximum at the density jump (position of the hydrodynamic discontinuity) and
we also note the presence of a cooling region (or Zel'dovich spike) in the temperature plot (b) at the position of the density jump, where
the gas temperature exceeds the post-shock temperature. The precursor measures approximately 0.02 cm and the gas opacity (which is averaged
over the entire frequency range) lies between $5 \times 10^{4}$ and $3 \times 10^{5}~\text{cm}^{2}~\text{g}^{-1}$ throughout (note that the
size of the precursor is measured between the shock position and the first point from the right hand side where the gas temperature exceeds
1.1 eV).

\begin{figure*}[!ht]
\centering
\includegraphics[scale=0.49]{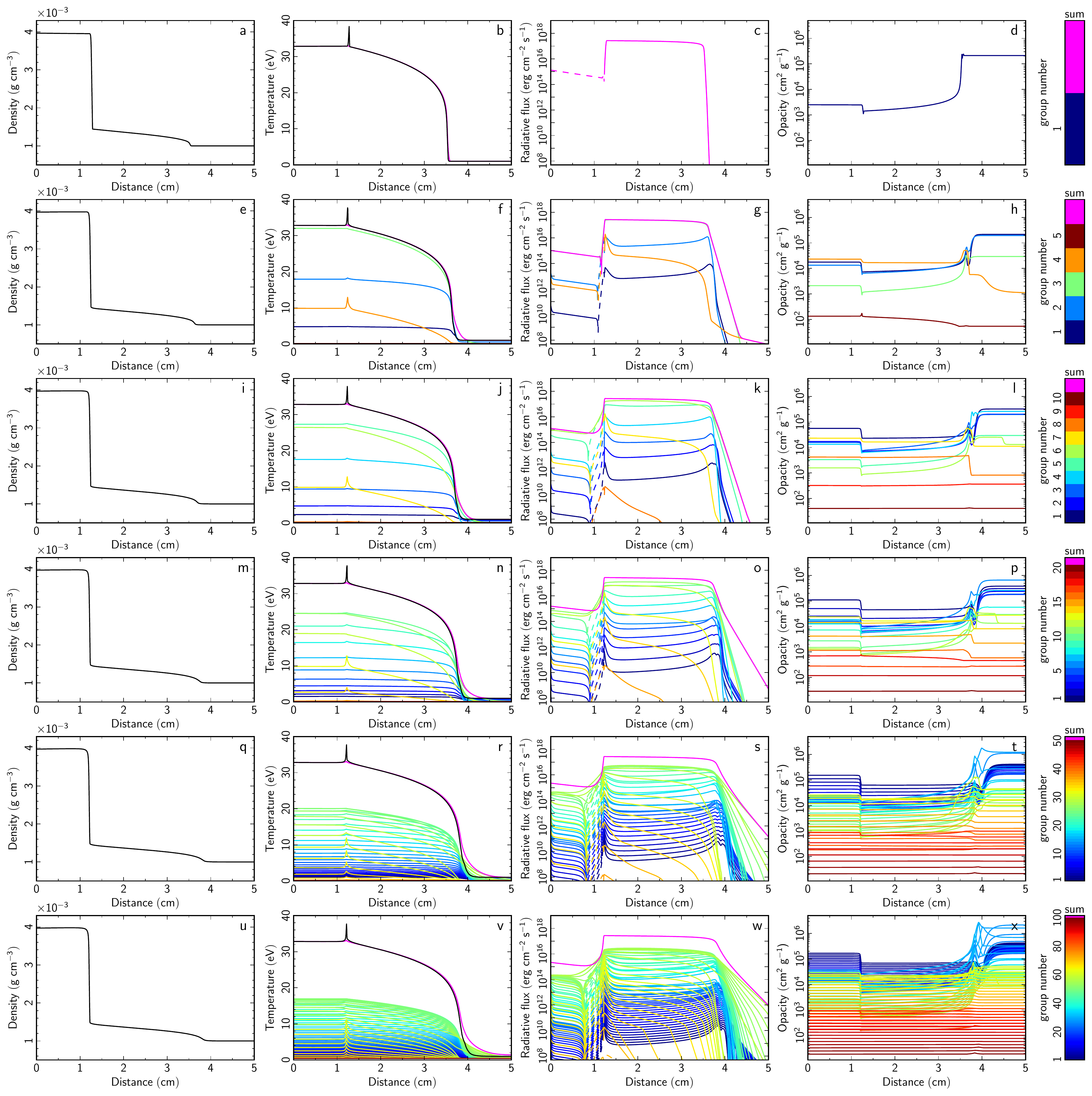}
\caption{Same as Fig.~\ref{fig:argon_sub_shock} for the supercritical shock case.}
\label{fig:argon_sup_shock}
\end{figure*}

The first multigroup simulation was performed using 5 frequency groups (\texttt{SUB005}), the results of which are shown in the second row.
The first major difference that emerges when compared to the grey simulation is that the extent of the radiative precursor has increased;
even though the gas temperature profile has not changed significantly, the radiative temperature curve extends much further ahead of the
discontinuity. This can be explained by the range of opacities observed in the different groups (see Fig.~\ref{fig:argon_sub_shock}h), and
in particular the opacity of group 3 (light green), whose radiative energy dominates inside the precursor, which is almost an order of
magnitude lower than the grey average. The grey opacity is biased towards a higher value which greatly affects the results, especially the
radiative flux which now shows a long tail extending towards the right end of the simulation box.

The results of the subsequent runs, using 10, 20, 50 and 100 groups, are shown in the lower rows. While runs \texttt{SUB005} and \texttt{SUB010}
are very similar to each other, the gas temperature profile has noticeably changed further in runs \texttt{SUB020} to \texttt{SUB100}, this being
most visible between 0.25 and 0.3 cm where the gas temperature is significantly altered by the radiation. The gas temperatures are overlayed
inside the same window in Fig.~\ref{fig:precursor_positions_and_temperatures}a for a clearer view. The position of
the density jump (or hydrodynamic shock) and the size of the precursor for all the simulations are listed in Table~\ref{tab:simulations_results}.
The precursor sizes keep increasing with the number of frequency groups used. A greater resolution in the frequency domain allows an accurate
treatment of the sharp slopes in the opacity curve (cf. Fig.~\ref{fig:kappanugr}), which affects
the amount of absorption ahead of the shock and defines the extent of the precusor. The extent of the precursor is plotted as a function of time
in Fig~\ref{fig:precursor_positions_and_temperatures}b. The influence of the number of groups described above is very clear on this plot.

\subsection{The supercritical case}\label{sec:sup_shock}

The results for the simulations of a supercritical radiative shock using $1-100$ frequency groups are shown
in Fig.~\ref{fig:argon_sup_shock}. The different rows and columns are identical to Fig.~\ref{fig:argon_sub_shock}.
The profiles are this time characteristic of a supercritical radiative shock; a diffusive radiative precursor strongly heats the gas ahead of
the shock, notably altering the gas density, the post-shock gas and radiative temperatures are identical to the pre-shock ones and the Zel'dovich
spike is clearly visible. As in the subcritical case, the size of the precusor increases with the number of frequency groups, growing by more than
20\% between 1 and 100 groups, as explicited in Table~\ref{tab:simulations_results} and illustrated in Figs~\ref{fig:precursor_positions_and_temperatures}c and 
\ref{fig:precursor_positions_and_temperatures}d. The fact that there is very little difference between the 50 and 100-group simulations indicate that we
have almost reached convergence of our results. The number of frequency groups used impacts the results less than in the subcritical case, since
most of the radiative precursor is in the diffusive regime where the grey approximation is deemed to be accurate.

These findings have important consequences on predictions made by numerical simulations on the structure of radiative shocks. Indeed,
due to the very different precursor sizes, studies of radiative shocks which make use of comparisons between observations and numerical
calculations will most probably be inaccurate if a grey radiative transfer model is used. One can compare the right column of
Fig.~\ref{fig:precursor_positions_and_temperatures} to the shock-precursor position diagrams found in the literature (see Fig.~3 in
\citealt{michaut2009} and Figs.~6 and 7 in \citealt{gonzalez2006} for example).

\subsection{Electron densities}\label{sec:electron_densities}

\begin{figure*}[!ht]
\centering
\includegraphics[scale=0.33]{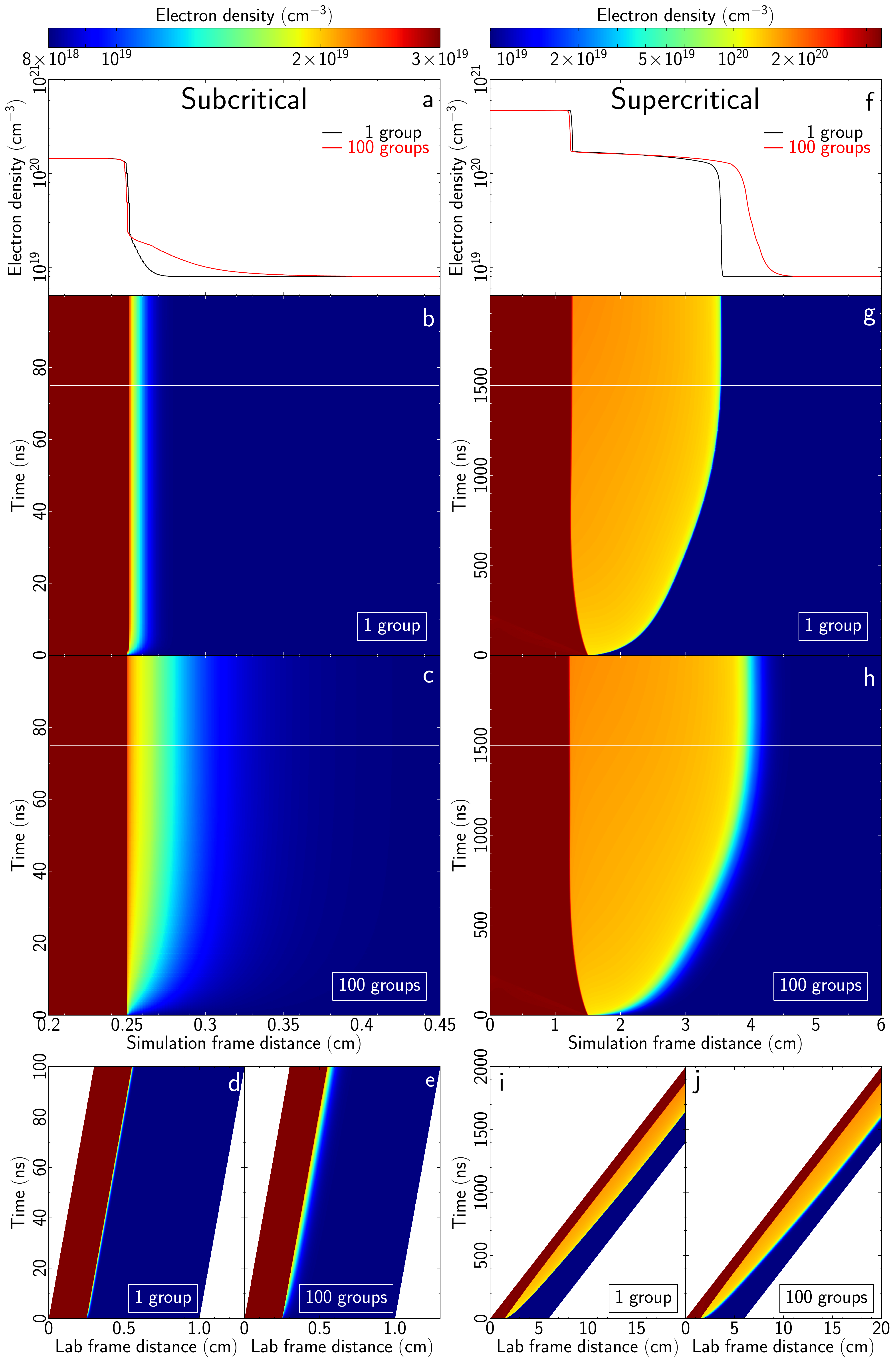}
\caption{Time evolution of the electron density ($N_{e}$) as a function of distance for the subcritical case (left column) and the supercritical
case (right column) using 1 and 100 frequency groups. Panels (a) and (f) show a slice through the 1 (black) and 100 (red) groups simulation rest
frame data. Panels (b) and (g) show the distribution of $N_{e}$ in the simulation frame for 1 group while panels (c) and (h) are for the 100-group
case. The bottom row displays the $N_{e}$ in the laboratory frame. The horizontal white line in panels (b), (c), (g) and (h) show the position at
which the slices were extracted. The bottom panels (d), (e), (i) and (j) show the $N_{e}$ distribution in the laboratory frame.}
\label{fig:electron_density}
\end{figure*}

In order to illustrate the differences between the grey and multigroup simulations further, we now consider the effects of our results on
the observables commonly obtained in radiative shock experiments. We compare in Fig.~\ref{fig:electron_density} the evolution of the electron density
$N_{e}$ as a function of time and distance in the grey and the 100-group simulations for the subcritical (left column) and the supercritical shocks
(right column). The middle panels (b), (c), (g) and (h) show the $N_{e}$ distribution in the simulation frame. The dark red region is the post-shock
final state with a dense and highly ionised medium, while the dark blue region is the pre-shock initial state. Between the two, the shock precursor
is clearly visible in yellow. The sharp transition between the precursor (yellow) and the post-shock state (red) represents the position of the
density jump as a function of time. The simulation frame panels are very useful in highlighting the differences between the grey and multigroup
simulations; the radiative precursor is unmistakably larger in the multigroup case. The top panels (a) and (f) show a slice extracted from the
simulation frame data, for an alternative view. We note that for the supercritical shock, the rightmost tip of the precursor is much sharper in the
grey than in the multigroup case. The supercritical simulation frame figures also reveal the slight displacement of the density jump towards the left,
as the whole structure of the radiative shock develops over time.

The bottom panels (d), (e), (i) and (j) show the $N_{e}$ distribution in the laboratory frame, and are reminiscent of Fig.~3 in \citet{michaut2009}.
These simulated laboratory-frame diagnostics show that the differences between the grey and multigroup simulations would be large enough to be
detected in experimental observations.

\subsection{Detection of adaptation zones around the Zel'dovich temperature spike}\label{sec:adaptation_zones}

\begin{figure*}[!ht]
\centering
\includegraphics[scale=0.44]{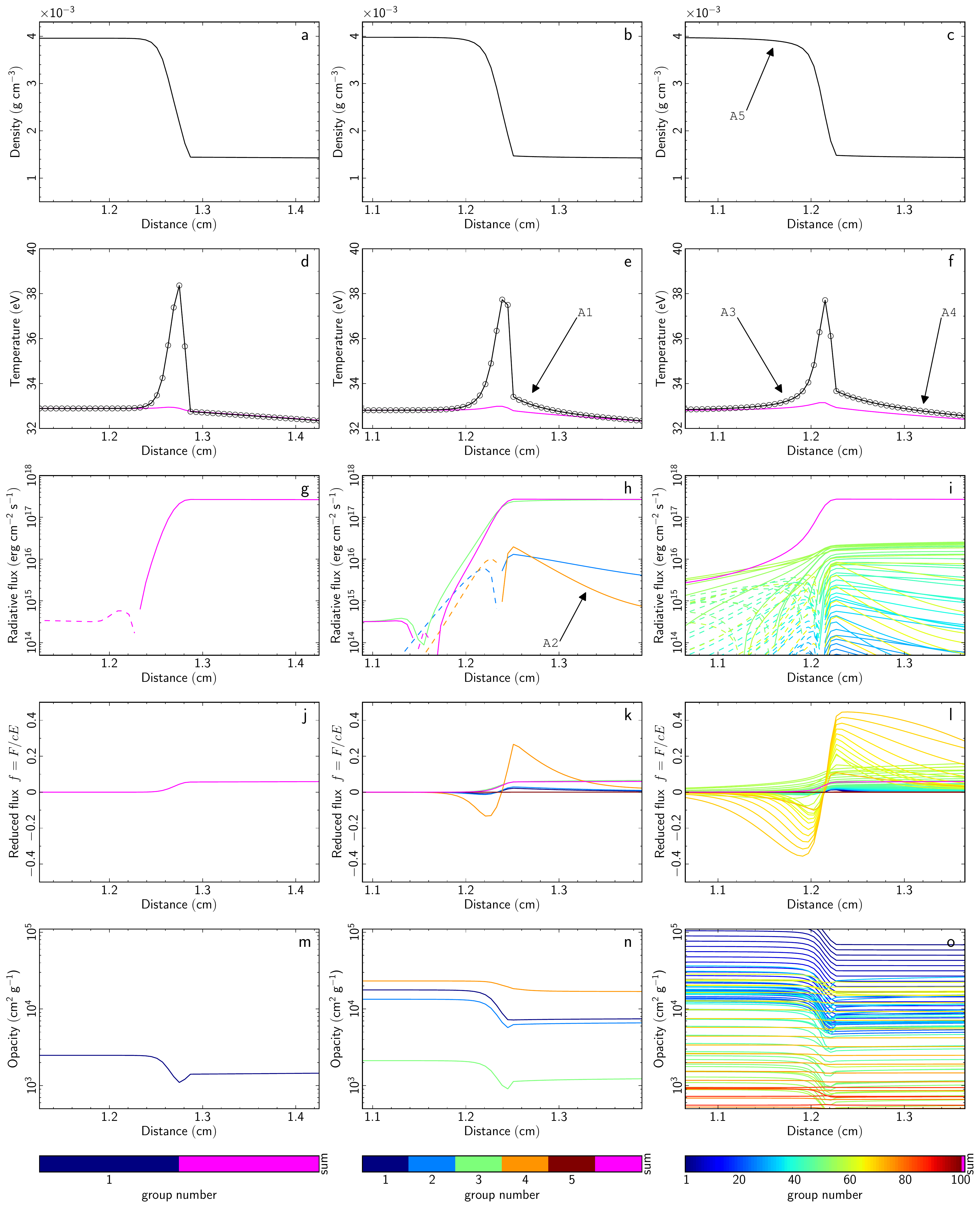}
\caption{A close-up view on the region around the Zel'dovich temperature spike for runs \texttt{SUP001} (left), \texttt{SUP005} (center) and
\texttt{SUP100} (right). From top to bottom: gas density, gas (black) and radiative (magenta) temperature, radiative flux, reduced flux and gas
opacity. In the temperature panels, the circles indicate the grid cells. As in the previous figures, the magenta curves represent the radiative
quantities summed over all groups and the dashed lines symbolise negative values of the radiative flux.}
\label{fig:argon_sup_shock_zoom}
\end{figure*}

We now turn to describe what is happening in the vicinity of the hydrodynamic discontinuity. Figure~\ref{fig:argon_sup_shock_zoom} shows a close-up
around the Zel'dovich spike for runs (from left to right) \texttt{SUP001}, \texttt{SUP005} and \texttt{SUP100}. The temperature profile of run
\texttt{SUP001} shows that the classic cooling layer structure, with a sharp edge on the right and a smooth cooling (or relaxation) region on the
left (as depicted in Fig.~\ref{fig:sup_shock_structure_simple}), is almost resolved (see Appendix~\ref{sec:appendixB} for a discussion on resolution).

However, the temperature profile of run \texttt{SUP005} shows a rather different structure, especially on the right hand side. There is a smooth
region of `over-temperature' to the right of the spike, spanning from $x = 1.26$ to 1.34 cm (indicated by arrow \texttt{A1}), where the gas temperature
$T_{-}$ is above the radiation temperature and higher than the final equilibrium temperature $T_{1}$. It progressively returns to the same value
as the radiation temperature as we move away from the spike towards the right. In a supercritical radiative shock, the pre-shock gas temperature
cannot in principle exceed the post-shock temperature \citep{zeldovich1967}. The solutions are converged in both cases and the radiative precursors
do not extend out to the right edge of the box. This effect must have another origin.

Looking at the radiative fluxes (panels g and h), we can see that in run \texttt{SUP001}, the flux shows a plateau right after the discontinuity. In
\texttt{SUP005}, the same is observed for the dominant group 3 (light green) but group 4 (orange) does not show a plateau to the right of the shock,
instead it decreases smoothly as we move away from the shock (see arrow \texttt{A2}). This indicates that the radiation in group 4 is being absorbed by the
material to the right of the shock. Indeed, the group 4 opacity of the pre-shock material (panel n) is about an order of magnitude higher than the
grey average in \texttt{SUP001} (panel m) where the radiation can escape `freely'.

It appears that the high opacity in group 4 causes the pre-shock material to absorb the radiation in that group only, and this consequently heats up
the gas a little more. It is like having a small precursor inside a precursor. This can be understood in terms of the following.
A small amount of incident radiation energy absorbed by gas per unit angle, time, area and frequency is equal to the specific intensity loss, i.e.
\begin{equation}\label{equ:dE_abs1}
dE_{\nu}^{\mathrm{abs}} = - dI_{\nu} ~ \cos \theta ~ dA ~ d\omega ~ dt ~ d\nu
\end{equation}
which for a constant opacity along direction $s$ becomes
\begin{equation}\label{equ:dE_abs2}
dE_{\nu}^{\mathrm{abs}} = \kappa_{\nu} ~ I_{\nu} ~ \cos \theta ~ dA ~ d\omega ~ dt ~ d\nu ~ ds ~~~.
\end{equation}

~\\~\\~\\~\\~\\ \noindent In one dimension, integrating over frequency and solid angle this reduces to
\begin{alignat}{2}\label{equ:E_abs1}
E^{\mathrm{abs}} & = \int_{0}^{\infty} \kappa_{\nu} \int_{-1}^{+1} \mu I_{\nu} d\mu ~ d\nu ~ dA ~ dt ~ ds\nonumber\\
                 & = \int_{0}^{\infty} \kappa_{\nu} F_{\nu} ~ d\nu ~ dA ~ dt ~ ds
\end{alignat}
where $\mu$ is the direction cosine. So the energy absorbed is proportional to $\int_{0}^{\infty} \kappa_{\nu} F_{\nu} ~ d\nu$. We now suppose
that we have two frequency groups, and to mimic the situation in run \texttt{SUP005}, we define the group quantities in the following way.
The two groups have the same width in the frequency dimension $\Delta \nu$, and we further assume the radiative flux and gas opacity to be constant
within each group, i.e. $\int_{0}^{\infty} \kappa_{\nu} F_{\nu} ~ d\nu = \sum_{g} \kappa_{g} F_{g} \Delta\nu$. The first group can be compared to group 3 in
\texttt{SUP005}; it dominates the radiative energy and flux, but has small reduced flux and opacity. On the other hand, the second group has a small
energy and flux, but a large opacity and reduced flux (similar to group 4 in \texttt{SUP005}). We choose
\begin{equation*}
\begin{array}{ccc}
\underline{\mathrm{Group~1}}: & ~ & \underline{\mathrm{Group~2}}:\\
~ & ~ & ~ \\
\begin{array}{lcl}
E_{1}      & = & E_{0}\\
\kappa_{1} & = & 0.1\\
f_{1}      & = & 0.05\\
F_{1}      & = & 0.05 c E_{0}
\end{array}
& ~ &
\begin{array}{lcl}
E_{2}      & = & E_{0} / 100\\
\kappa_{2} & = & 1.0\\
f_{2}      & = & 0.5\\
F_{2}      & = & 0.005 c E_{0}
\end{array}
\end{array}
\end{equation*}
The integral over the frequency range in equation (\ref{equ:E_abs1}) now becomes a sum over the two groups and we have
\begin{alignat}{2}\label{equ:E_abs2}
E^{\mathrm{abs}} & \propto [ \kappa_{1} F_{1} + \kappa_{2} F_{2} ] \Delta\nu\nonumber\\
                 & \propto [ 5 \times 10^{-3} c E_{0} + 5 \times 10^{-3} c E_{0} ] \Delta\nu
\end{alignat}
which is twice as much as what is found by considering only the dominant group 1. This shows that a group with very little radiative energy and
radiative flux can contribute significantly to the total energy absorbed if it has a large reduced flux and opacity. The fact that
$E_{1}/E_{2} = 100$ means that including group 2 does not change the total radiative temperature, but does change the amount of energy absorbed
by the gas, and the gas and radiative temperatures can hence be decoupled.

The morphology of the Zel'dovich spike changes again for run \texttt{SUP100} (Fig.~\ref{fig:argon_sup_shock_zoom}f). The decoupled regions both
to the left and the right of the hydrodynamic discontinuity are wider than for \texttt{SUP005} (see arrows \texttt{A3} and \texttt{A4}). The
radiative flux diagram (i) this time shows more components with a noticeable downward trend and large reduced fluxes. The range of opacities in
the numerous groups mean that the radiation in different groups are absorbed at different rates, which yields a wider relaxation region to the
right of the shock. The same effect is also observed to the left of the spike where the gas and radiation temperatures are also decoupled (with
negative fluxes travelling from right to left).

The structure observed here is what is described as an adaptation zone by \citet{drake2007a,drake2007b}; a region across which the influence of
radiation from the cooling layer on the shock structure fades away and where the temperature and other gas quantities make their final small
adjustments in order to reach their final steady-state values. Figure 1 in \citet{drake2007a} actually depicts exactly the situation we observe
here. There are two adaptation zones on each side of the cooling layer (i.e. the Zel'dovich spike) where the gas temperature is higher than the
final state $T_{1}$ downstream of the density discontinuity and higher than the precursor temperature $T_{-}$ upstream. This is precisely what
our simulation results show for 10 frequency groups and above. \citet{drake2007a} actually depicts the pre-shock temperature just before the
discontinuity as below or equal to the final state $T_{1}$, but he does mention that ``ongoing numerical work by John Castor suggests that the
temperature inside the adaptation zone, at the actual density jump, may be pulled up above [$T_{\mathrm{1}}$]''.

The extent of the region inside which the radiative flux from the cooling layer still has an effect on the surrounding gas will inevitably depend
on the opacity of the gas, which explains the different observed sizes for the adaptation zones as we vary the number of groups. The density is
also pictured in Drake's papers as being slightly lower than the final state $\rho_{1}$ inside the downstream adaptation zone and higher than the
leftmost precursor value $\rho_{-}$ in the upstream zone. This is also the case in our results, as shown by arrow \texttt{A5} in
Fig.~\ref{fig:argon_sup_shock_zoom}c.

\citet{drake2007b} provide an analytical estimate of the width of the spike for a given shock strength, and when applied to our simulation setup,
it predicts that the spike should be narrower by a factor $\sim30$. \citet{mcclarren2010} also mention that the $M_{1}$ model, even though it
describes the total energy flows correctly, might not perform satisfactorily in the vicinity of the cooling layer, but this has yet to be
investigated. It would appear from the analytical estimates that we might not be resolving the real `physical' Zel'dovich spike, but the
discussion in Appendix~\ref{sec:appendixB} shows that we are sufficiently resolving the spike (for the purpose of this study) resulting from our
numerical model. We also demonstrate that our observation of adaptation zones being absent from grey simulations while being clearly detected in
multigroup simulations is not resolution dependent. While we have to acknowledge that we might not be resolving the true width of the spike, as
we do not attempt to make any comparisons with experiments, we believe that our results are still legitimate.

\begin{figure}[H]
\centering
\includegraphics[scale=0.65]{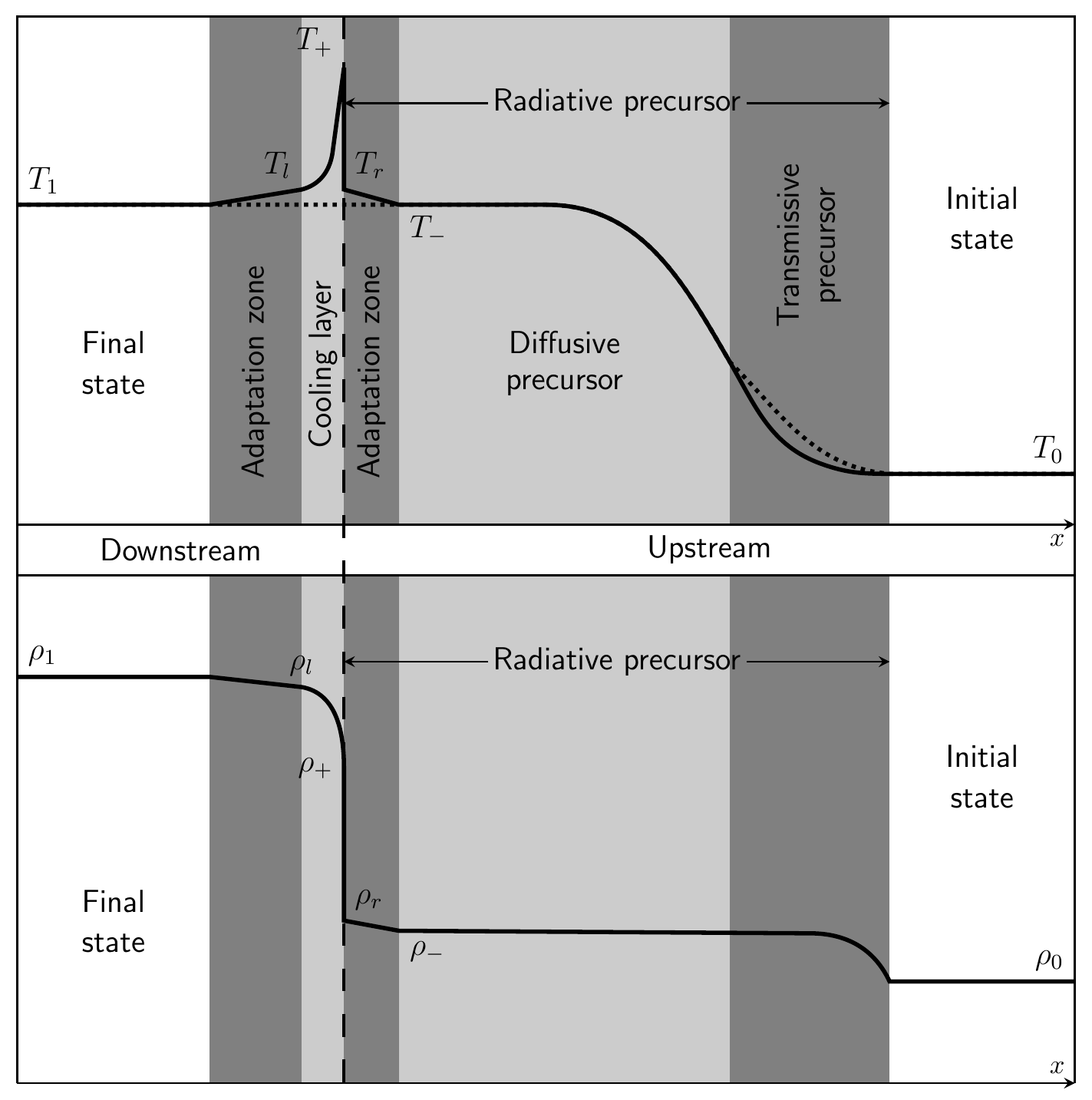}
\caption{Structure of a supercritical radiative shock \citep[adapted from][]{drake2007a}. The direction of the gas flow is from right to left in the
frame where the shock is at rest. Top panel: gas (solid) and radiative (dotted) temperature as a function of distance. Bottom panel: gas density as a
function of distance. The position of the temperature and density jumps is marked by the vertical dashed line. The relative sizes of the layers are
for illustration purposes only.}
\label{fig:sup_shock_structure_complex}
\end{figure}

Finally, the presence of these adaptation zones in our simulations forces us to revise our depiction of a radiative shock structure and adopt a more
up-to-date description (see Fig.~\ref{fig:sup_shock_structure_complex}). The preshock gas is heated by the radiative precursor to a temperature
$T_{-} \approx T_{1}$, and increases slightly to a temperature $T_{r}$ as we cross the right adaptation zone through radiative heating. The shock
compression at the density jump (discontinuity) heats it further to a temperature $T_{+}$ which is higher than the final post-shock equilibrium state
$T_{1}$. The gas then cools down inside the cooling layer to reach the intermediate post-shock state $T_{l}$ by radiating the excess energy away. The
final small adjustments are made across the left adaptation zone to reach the final post-shock state $T_{1}$ \citep[see also][]{drake2007b}. The
radiative temperature (dotted line) is equal to the gas temperature for the most part, except that it remains constant across the cooling  layer 
and the adaptation zones (decoupled from the gas), and is higher than the gas temperature inside the transmissive precursor. The pressure gradient 
at the head of the precursor, through the conservation of mass (\ref{eq:cons_mass}) and momentum (\ref{eq:cons_mom}), causes the velocity to decrease
and the density to increase to a value $\rho_{-}$ ahead of the discontinuity. Since the gas temperature is close to being constant in the diffusive
part of the precursor, there is no more pressure gradient and the density reaches a plateau value $\rho_{-}$ ahead of the discontinuity. A small
compression from $\rho_{-}$ to $\rho_{r}$ occurs as we cross the right adaptation zone, then followed by the sharp density jump of the shock from
$\rho_{-}$ to $\rho_{+}$ which takes place on the gas viscous scale. The density then rises rapidly to $\rho_{l}$ inside the cooling layer through
strong contraction of the radiating gas. Lastly, the density slowly reaches the final state $\rho_{1}$ across the left adaptation zone.

\subsection{Hugoniot curves}\label{sec:hugoniot}

\begin{figure*}[!ht]
\centering
\includegraphics[scale=0.42]{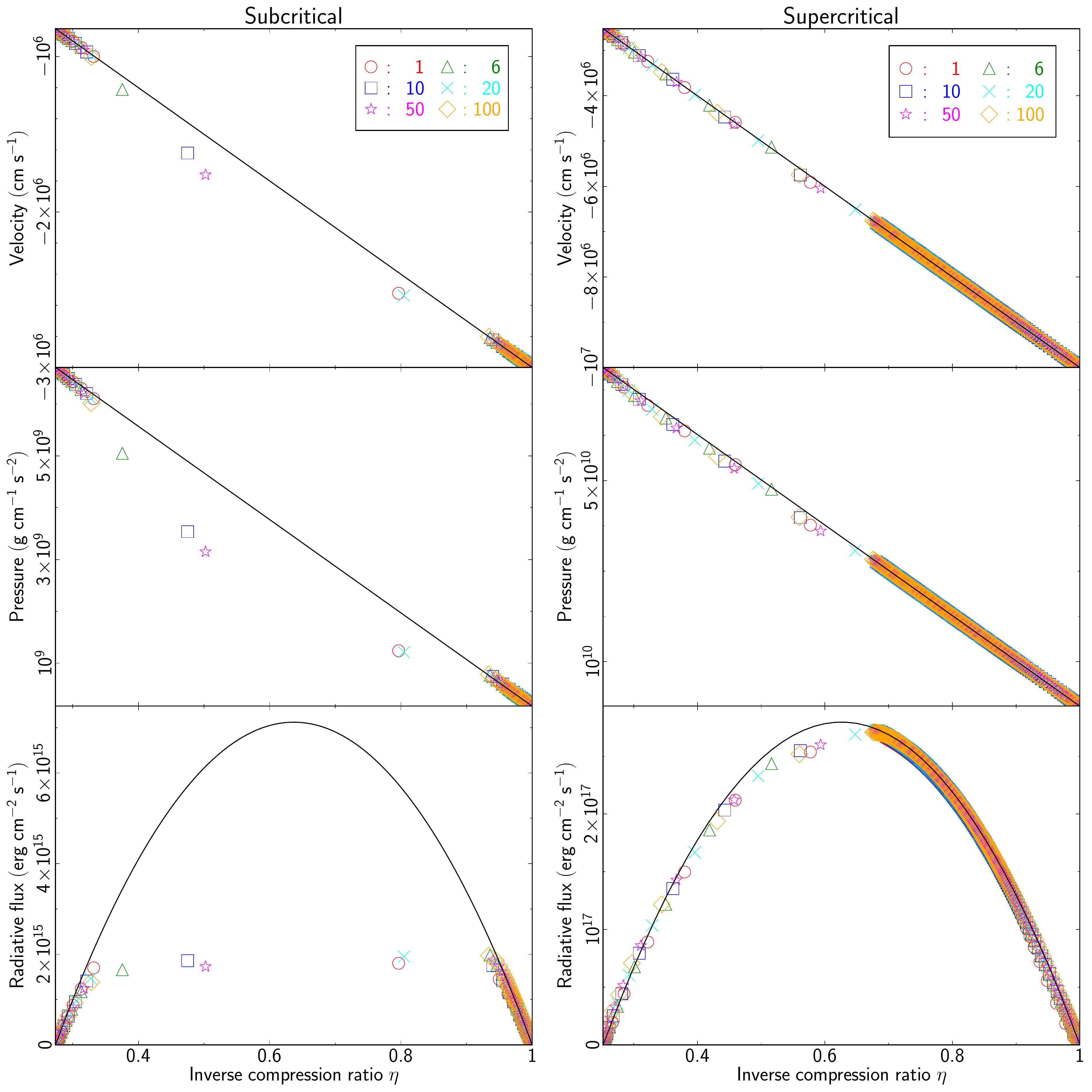}
\caption{Hugoniot curves for the radiative shock simulations using $1-100$ frequency groups (see legend in the top right corner for the meaning of
the symbols): gas velocity (top), pressure (middle) and radiative flux (bottom) as a function of inverse compression ratio in the case of the
subcritical shock (left column) and the supercritical shock (right column). The analytical solutions are overlayed for comparison (black solid line).
The simulation points not lying on the analytical solutions are all within the shock transition.}
\label{fig:hugoniot}
\end{figure*}

In this section we look at the Hugoniot curves for gas velocity, pressure and radiative flux in our simulations. The Hugoniot curves are analytical
predictions for the state of gas quantities as a function of the inverse compression ratio $\eta = \rho_{0}/\rho$ for which conservation of mass,
momentum and energy hold \citep[see][for example]{zeldovich1967,mihalas1984}. Figure~\ref{fig:hugoniot} shows (from top to bottom) the gas velocity,
pressure and radiative flux as a function of inverse compression ratio for every grid point in our subcritical (left) and supercritical (right)
simulations, as well as the analytical solutions (solid black line) which are
\begin{align}
u(\eta) & = - u_{0} \eta \label{eq:hug_vel}\\
p(\eta) & = \rho_{0} u_{0}^{2} (1-\eta) + p_{0} \label{eq:hug_pres}\\
F(\eta) & = \frac{\rho_{0}u_{0}^{3}}{2}\left(\frac{2\gamma}{\gamma-1}\eta - \frac{\gamma+1}{\gamma-1}\eta^{2} - 1 \right) - \frac{\gamma p_{0}u_{0}}{\gamma-1}(1-\eta) + F_{0} \label{eq:hug_Fray}
\end{align}
where $\gamma$ is the ratio of specific heats and the radiative flux $F$ is expressed in the laboratory frame \citep[note that we have converted our
radiative quantities to the laboratory frame using equations 91.16 and 91.17 in][]{mihalas1984}.

This illustrates well how our numerical scheme preserves the important physics of conservation of mass, momentum and energy. Any simulation of
shocks which has points lying away from these analytical curves does not conserve these fundamental quantities. The points which do not lie on the
analytical curves in Fig.~\ref{fig:hugoniot} are all in the shock transition region which is inevitably spread over a minimum number of cells for a
Godunov method and do not strictly match the analytical solution \citep[see][for example]{sincell1999,drake2007b}. This is mostly visible in the
subcritical case where the radiative flux follows a constant value transition while the gas velocity and pressure follow an approximate Hugoniot
curve through the transition to connect the upstream and downstream states. Note that only a single cell for each simulation does not lie on the
analytical prediction, illustrating the fact that the transition is spread over only one or two grid cells. We take the opportunity here to point
out that the comoving frame formalism, which has been criticized in several works for not rigorously conserving the total energy
\citep[see][for example]{mihalas1982,krumholz2007}, appears here to perform very well in conserving the fundamental quantities.

\section{Conclusions}\label{sec:conclusion}

We have performed simulations of stationary radiative shocks in Ar gas using a multigroup radiation hydrodynamics scheme. Gas opacities
depending on temperature, density and frequency were used in the equations of radiation hydrodynamics to achieve convincing results. The simulations
reproduced all the detailed structure of a radiative shock, including the radiative precursor, the cooling layer and even the adaptation zones
connecting the cooling layer to the final downstream state and the precursor. Our results show that grey simulations produce very different results
compared to multigroup ones, and that frequency-independent calculations are not deemed an accurate description of the problem. Indeed, multigroup
simulations showed increases by a factor of four in precursor size in the subcritical case, and an increase of 20\% in the supercritical case. The
simulations with 5 to 100 groups also show increasing precursor sizes, with a suspected convergence of results for 50 groups and above. Multigroup
effects were also seen to be important in the vicinity of the cooling layer, where adaptation zones absent from grey simulations were clearly
detected.

We have to acknowledge that several caveats need to be taken into account when considering the results reported in this work. Firstly, the exact
sizes of the radiative precursors are not entirely correct since higher resolution simulations reported in Appendix~\ref{sec:appendixB} yield slightly
different results (typical differences are of the order of $5-10$\%), only the relative increase in precursor sizes between the different simulations
are of notorious relevance. Secondly, even though our results are numerically quantitatively converged in the proximity of the Zel'dovich spike, it is
not clear whether the $M_{1}$ model performs accurately enough on such small scales. We remind the reader that this work is not an attempt at directly
comparing numerical simulations to experiments, merely a study of the effects of frequency dependence on the results of radiation hydrodynamic
calculations.

Nevertherless, the findings presented still have important consequences on predictions made by numerical simulations on the structure of radiative
shocks. Indeed, studies of astrophysical (accretion processes, supernova remnants, jets, etc\dots) or laboratory radiative shocks will most probably
be inaccurate if a grey radiative transfer model is used. The impact can not only be large when looking at the radiative precursor sizes, but also
on the total energy budget, determining the amount of energy converted into radiation and absorbed by pre-shock material.

It is difficult to compare the results of this work with experimental data directly since our idealised setup of stationary radiative shocks is very
far from the situation in laboratories where laser-driven radiative shocks travel through gas chambers and it is often unclear if they ever reach a
stationary state. Realistic calculations using a piston-like shock-driving boundary, as well as a more sophisticated equation of state, will be more
appropriate for conducting detailed modelling of laboratory experiments.

\section*{Acknowledgements}\label{sec:acknow}

The research leading to these results has received funding from the European Research Council under the European Community's Seventh Framework
Programme (FP7/2007-2013 Grant Agreement no. 247060). The authors also greatfully acknowledge support from grant ANR-06-CIS6-009-01 for the programme
SiNeRGHy. The authors would also like to thank M. Busquet for fruitful discussions which have yielded vast improvements of this paper and J.~P. Gauthier
for the Ar opacity tables. Finally, the authors would like to thank the reviewers for very useful comments which have helped greatly in improving the
robustness and credibility of the paper.

\appendix

\section{Using the correct opacity average}\label{sec:appendixA}

In the RHD equations (\ref{eq:cons_mom})--(\ref{eq:cons_Fr}), it is challenging to compute the radiative energy and flux-weighted mean opacities
$\kappa_{E}$ and $\kappa_{F}$ (note that $\sigma = \kappa \rho$), since the quantities $E_{\nu}$ and $F_{\nu}$ are not necessarily known at \emph{all}
wavelengths. The choice of approximation for these quantities, which are crucial to the RHD calculations, is not trivial \citep[see][p. 88+ for a
discussion]{pomraning1973}. Common practise is to set $\kappa_{E} = \kappa_{P}$ and $\kappa_{F} = \kappa_{R}$ where $\kappa_{P}$ and $\kappa_{R}$ are
the Planck and Rosseland mean opacities, respectively \citep[see][for example]{larsen1994,offner2009}. However, for the $M_{1}$ model, numerical
stability is immensely improved if an identical value for $\kappa_{E}$ and $\kappa_{F}$ is used, which prompts us to use an average value
$\kappa_{\text{av}} = \kappa_{E} = \kappa_{F}$.

\begin{figure}[H]
\centering
\includegraphics[scale=0.4]{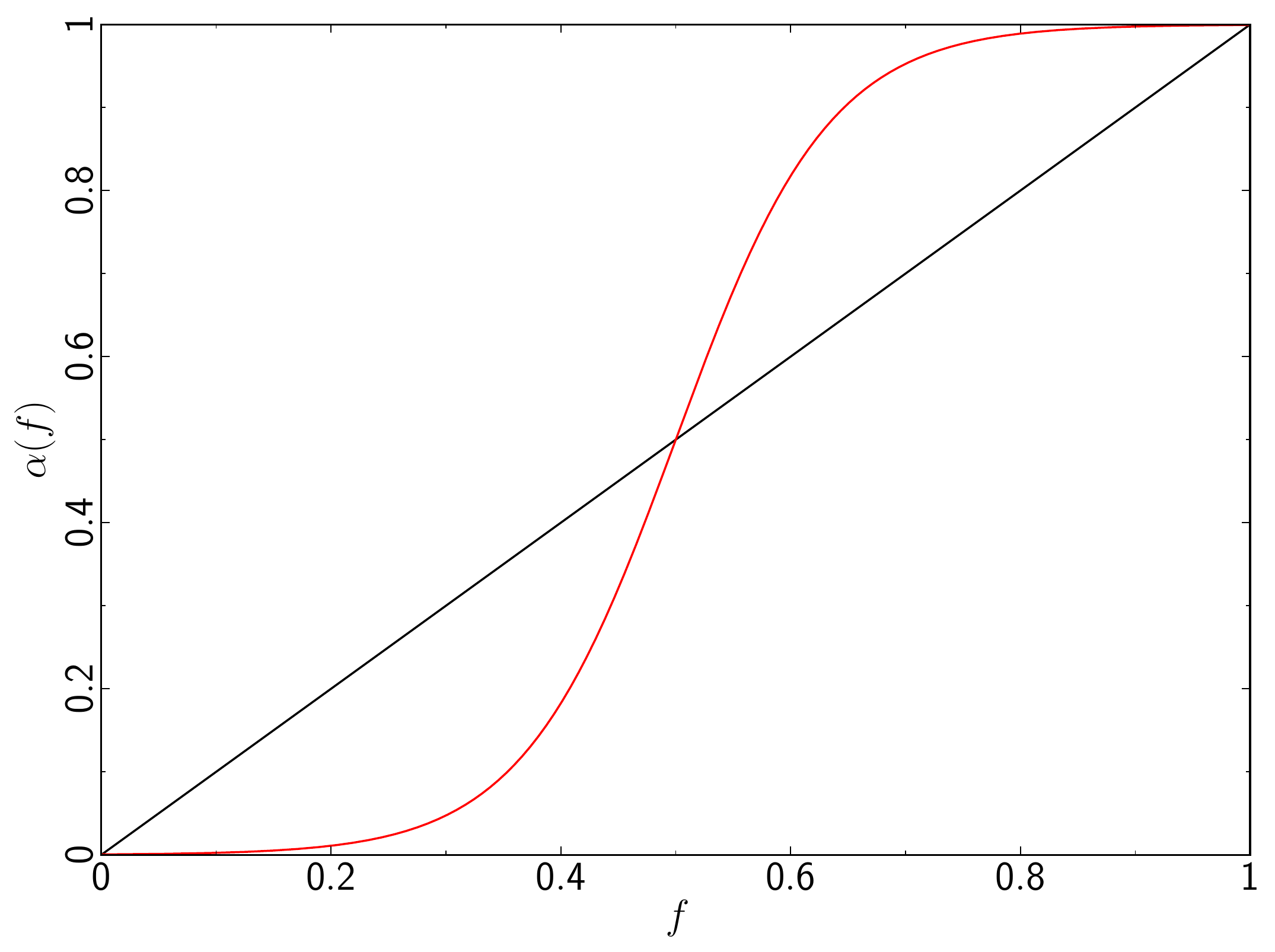}
\caption{$\alpha$ parameter as a function of reduced flux $f$: the black solid line is the simple $\alpha = f$ while the red line is given by equation
(\ref{eq:alpha}).}
\label{fig:alpha}
\end{figure}

The Planck and Rosseland means are applicable to different regimes. In the diffusion limit, the opacity should be equal to the Rosseland mean, while
the Planck mean is appropriate in the free-streaming limit. We need to make sure that we recover this behaviour with our average opacity.
\citet{sampson1965} proposed an average which varies between $\kappa_{P}$ and $\kappa_{R}$ depending on the optical depth. Here, we use the reduced
radiative flux $f = \| \mathbf{F} \| / c E$ as a measure of the diffusivity (optically thick) or transmissivity (optically thin) of the flow. We
define a parameter $\alpha(f)$ which varies between zero and one according to $f$ which then allows us to write the average opacity
\begin{equation}\label{eq:kappa_av}
\kappa_{\text{av}} = (1 - \alpha) \kappa_{R} + \alpha \kappa_{P} .
\end{equation}
In order to recover the diffusion and free-streaming limits, $\alpha$ needs to have the following properties: $\alpha \rightarrow 0$ when $f \rightarrow 0$
and $\alpha \rightarrow 1$ when $f \rightarrow 1$. The simplest formula with these properties is just the linear function $\alpha = f$. However, we argue
that the $\alpha$ parameter is meant to represent the transition from a regime dominated by diffusion to a radiation-dominated regime. In this sense, one
expects the transition from one to the other to be rather rapid, as opposed to a smooth linear averaging between the $\kappa_{P}$ and $\kappa_{R}$ values.
We thus propose a different formula for $\alpha$ which will better reproduce this behaviour. After some experimenting, we finally chose
\begin{equation}\label{eq:alpha}
\alpha_{s} = \frac{1}{e^{-15(f - 1/2)} + 1}
\end{equation}
which is plotted in Fig.~\ref{fig:alpha} (red) alongside the simpler $\alpha = f$ (black). We would like to point out here that we have no physical
explanation for equation (\ref{eq:alpha}), it is simply an \textit{ad-hoc} choice of a function with the correct properties. However, we can justify
our choice by testing the method in a simple case.

\begin{figure*}[!ht]
\centering
\includegraphics[scale=0.4]{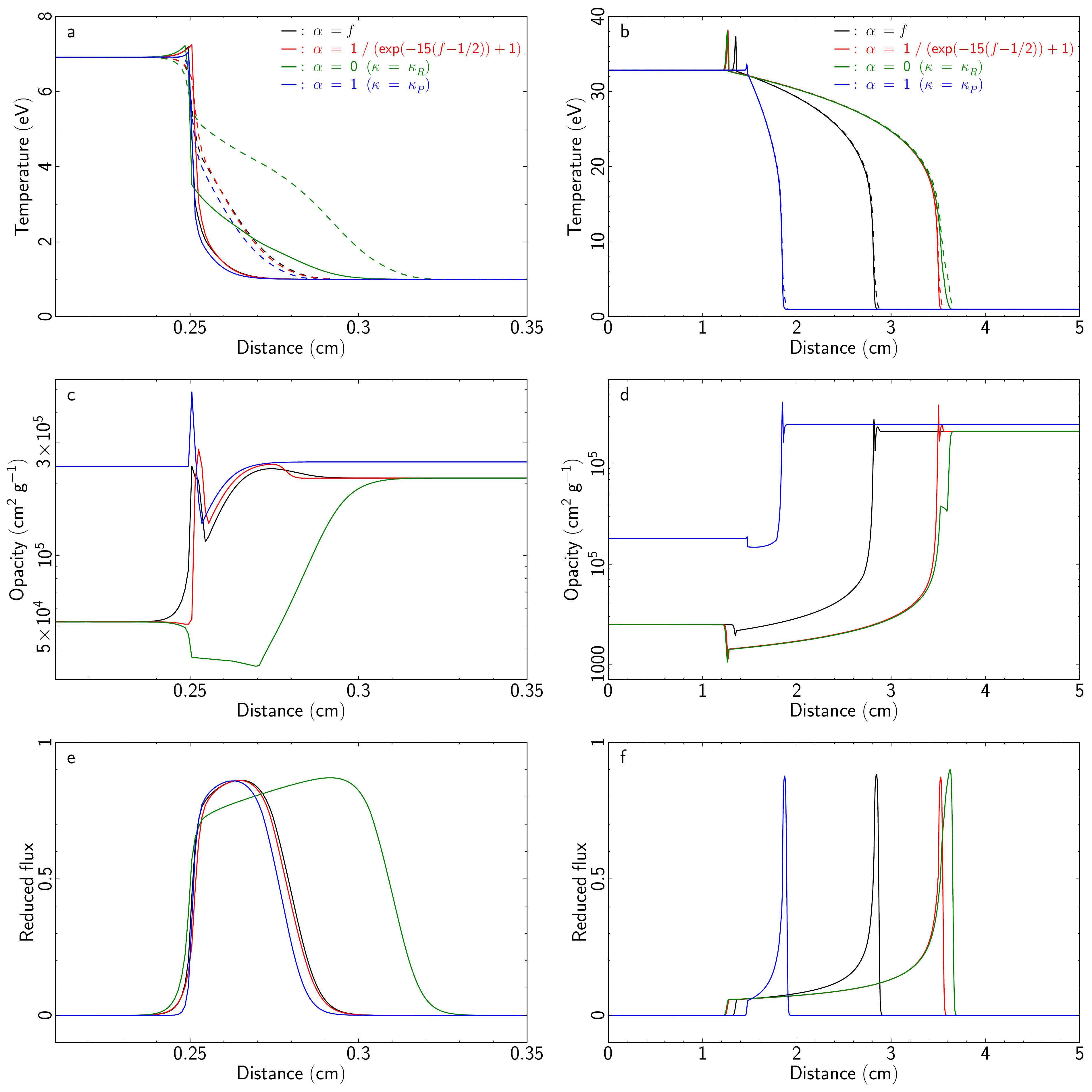}
\caption{Comparison of the different opacity averaging functions. The left and right columns show the results for the subcritical and supercritical
radiative shocks, respectively. The top row displays the gas (solid) and radiative (dashed) temperatures, the middle row is the gas opacity and the
bottom row shows the reduced flux. The colour-coding is as follows: black is for $\alpha = f$, red is for $\alpha$ given by equation (\ref{eq:alpha}),
green is for $\kappa_{\text{av}} = \kappa_{R}$ and blue is for $\kappa_{\text{av}} = \kappa_{P}$.}
\label{fig:kappa_av_shocks}
\end{figure*}

We performed eight simulations with a single frequency group; four calculations of a subcritical radiative shock and four of a supercritical shock. In
each case, the first was carried out using $\alpha = f$, the second using $\alpha = \alpha_{s}$, the third using $\alpha = 0$ (which corresponds to
$\kappa_{\text{av}} = \kappa_{R}$) and the fourth using $\alpha = 1$ ($\kappa_{\text{av}} = \kappa_{P}$). We show the results in
Fig.~\ref{fig:kappa_av_shocks}. In the case of the subcritical radiative shock (left column), we see that the average opacities (black and red) are
equal to the Rosseland mean far away from the discontinuity (on both sides). They then become close to the Planck average in the intermediate region
(between 0.25 and 0.28 cm) where the reduced flux is large. The temperature plot shows that the results for both the averaging schemes are very close
to the blue curve (Planck mean) which is the desired result in this optically thin regime. For the supercritical shock (right column), the average
opacity should produce results which resemble the green curves, which are appropriate in this diffusive regime (only a small region of the grid has a
large $f$). This time, the simple $\alpha = f$ approximation shows its limitations with a precursor size about two thirds of the size of the one
observed in the $\kappa_{\text{av}} = \kappa_{R}$ case. On the other hand, the $\alpha = \alpha_{s}$ model performs much better, producing results very
similar to the $\kappa_{\text{av}} = \kappa_{R}$ simulation. We thus conclude that the expression given in equation (\ref{eq:alpha}) is an effective
model to simulate problems in both diffusive and free-streaming limits.

In a final note, we wish to reiterate that the inaccuracies which arise from the approximation of setting $\kappa_{E} = \kappa_{F} = \kappa_{\text{av}}$
are reduced as the number of frequency groups used increases, since in the limit of infinite frequency resolution, all of these quantities simply reduce
to $\kappa_{\nu}$. This approximation is thus less crude in a multigroup model than in a grey model.

\section{A comment on spatial resolution}\label{sec:appendixB}

\begin{figure*}[!ht]
\centering
\includegraphics[scale=0.4]{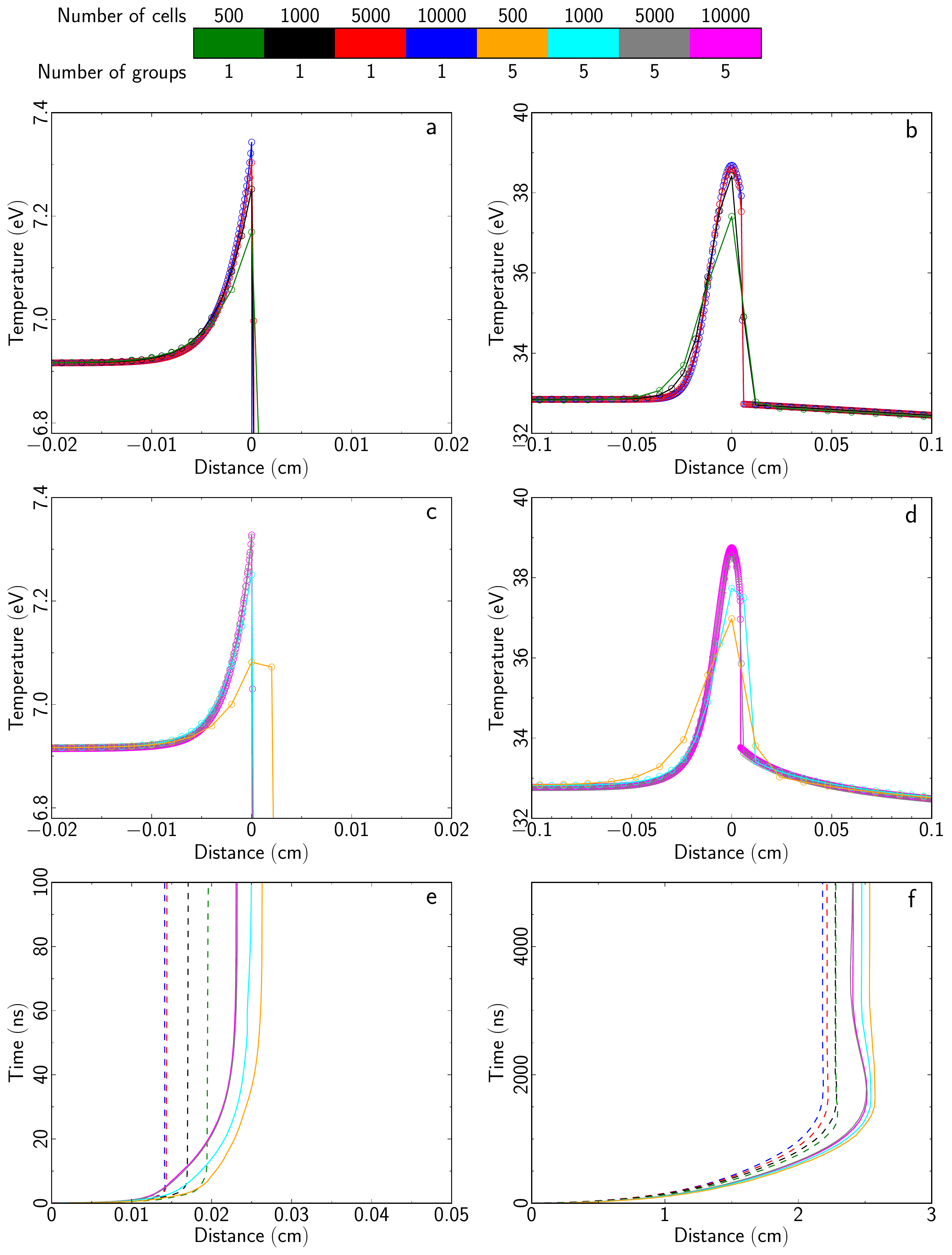}
\caption{Comparison between the effects of grid resolution and number of frequency groups. Top row: temperature profiles in the vicinity of the
cooling layer for 1 frequency group using different spatial resolutions in the subcritical (a) and supercritical (b) cases (see color key at the
top of the figure). Middle row: same as the top row but using 5 frequency groups. Bottom row: size of precursor as a function of time for the
1-group (dashed) and 5-groups (solid) simulations in the subcritical (e) and supercritical (f) cases.}
\label{fig:resolution}
\end{figure*}

In Fig.~\ref{fig:argon_sup_shock_zoom} it would appear that with our spatial resolution of 1000 grid cells, we do not fully resolve the very narrow
Zel'dovich spike. For completeness, we report in this section a short spatial resolution study to confirm that our results concerning the influence of
the number of frequency groups on the size of the radiative precusors and adaptation zones on each side of the spike are robust.
Figure~\ref{fig:resolution} compares the effects of grid resolution and number of frequency groups on the shock structures. The top row shows the
temperature profiles in the vicinity of the cooling layer for 1 frequency group using 500, 1000, 5000 and 10,000 cells in the subcritical (a) and
supercritical (b) cases (see color key at the top of the figure). In the subcritical case, the different resolutions appear to yield similar results.
For the supercritical runs, the spike appears well resolved in the high resolution simulations. It looks thinner than in the lower resolution
calculations and displays a (smooth) maximum to the left of the discontinuity, a structure that much resembles the results of \citet{lowrie2008}.
At a Mach number of $\sim 19$, the spike structure fits well between their depictions of shocks at $\mathcal{M}=3$ (Fig.~10) and $\mathcal{M}=27$
(Fig.~13). A strong convergence of results is observed between 5000 and 10,000 cells.

The middle row shows again the temperature profiles of the cooling layer for different resolutions but this time using 5 frequency groups. Panel (c)
reveals that 500 cells is most probably too few to resolve the cooling layer. Both panels also show convergence between 5000 and 10,000 cells.
More importantly, panels (b) and (d) demonstrate that our overall conclusions on the presence of adaptation zones in the multigroup simulations
remain. Even though the relaxation region to the right of the spike is more pronounced in the 5000-cell simulation (grey), the results from the
1000-cell simulations (cyan) agree qualitatively; an adaptation zone is absent from the grey simulations but is detected in the multigroup simulations. 

Analytical estimates of the width of the spike from \citet{drake2007b} suggest that the real physical spike for the same values of shock velocity and
initial state density might in fact be much narrower (by a factor of $\sim30$). Nevertherless, the fact remains that we have converged spatially on
the structure of the cooling layer given by our $M_{1}$ model, which is what matters for the present study. In addition, analytical estimates also
make use of approximations and the \emph{true} width of the spike is probably not known accurately.

The bottom row displays the size of the radiative precursors as a function of time for the different simulations. To better distinguish the separate
runs, we have plotted the simulations using 1 group with dashed lines and the simulations using 5 groups with solid lines (the colours remain the
same as in the other panels). These plots reveal that the spatial resolution also affects the total size of the precursor (here again the results
have converged for 5000 cells and above). However, as we do not make any direct comparisons with experiments or observations throughout this work but
are only interested in the relative differences in precursor sizes, we argue that our conclusions regarding the increase in precursor size as a
function of number of frequency groups remain qualitatively correct. Moreover, panels (e) and (f) show quite clearly that the difference in precursor
size due to a change in number of groups is larger than the difference observed from a change in number of cells. In addition, Fig.~\ref{fig:resolution}
only shows the results using 1 and 5 groups, and the size of the precursor is seen to continue increasing all the way up to 100 groups (see
Fig.~\ref{fig:precursor_positions_and_temperatures}). In the case of the subcritical shock, differences between runs with 1000 and 10,000 cells are
$\sim 20$\% and $\sim 8$\% for the 1-group and 5-groups simulations, respectively, while differences between the 1-group and 100-groups (1000 cells)
simulations are higher than 400\% (see Table~\ref{tab:simulations_results}). As for the supercritical, resolution alters the precursor sizes by only
$2-4$\% while frequency groups have an effect of the order of 20\%. Finally, the relative differences in precursor size between 1- and 5-group
simulations for a given resolution remain approximately constant. It was not possible for us to run a 100-group simulation using 5000 cells on a
realistic timescale, but we believe that the results of the 1000-cell simulations are qualitatively robust.

\end{multicols}

\end{document}